\documentclass{pasa}

\usepackage{graphicx}

\newcommand{\changed}[1]{ {
#1} }

\title[SPICA - a large cryogenic infrared space telescope]{SPICA - a large cryogenic infrared space telescope \\  {\em\LARGE Unveiling the obscured Universe}}

\author[P.R. Roelfsema et al.]{P.R. Roelfsema$^{1,2}$\thanks{P.R.Roelfsema@sron.nl},
          H. Shibai$^{3,4}$,
          L. Armus$^{5}$,
         \changed{ D. Arrazola$^{26}$,}
          M. Audard$^{6}$,
          M.D. Audley$^{1}$,
          C.M. Bradford$^{7,8}$,
          \changed{I. Charles$^{28}$,}
          \changed{P. Dieleman$^{1}$,}
          Y. Doi$^{9}$,
         \changed{ L. Duband$^{28}$,
          M. Eggens$^{1}$,
          J. Evers$^{1}$,
          I. Funaki$^{4}$,
          J.R. Gao$^{1}$,}
          M. Giard$^{10}$,      
         \changed{A. di Giorgio$^{11}$
          L.M. Gonz\'alez Fern\'andez$^{26}$,}
          M. Griffin$^{12}$, 
          F.P. Helmich$^{1,2}$,
         \changed{ R, Hijmering$^{1}$,
          R. Huisman$^{1}$,
          D. Ishihara$^{13}$,
          N. Isobe$^{4}$,}
          B. Jackson$^{1}$,
          \changed{H. Jacobs$^{1}$,
          W. Jellema$^{1}$,}
          I. Kamp$^{2}$, 
          H. Kaneda$^{13}$,
          \changed{M. Kawada$^{4}$,
          F. Kemper$^{14}$,}
          F. Kerschbaum$^{15}$,
          \changed{P. Khosropanah$^{1}$,}
          K. Kohno$^{16}$,
          \changed{P.P. Kooijman$^{1}$,}
          O. Krause$^{17}$,
          \changed{J. van der Kuur$^{1}$,
          J. Kwon$^{4}$,
          W.M. Laauwen$^{1}$,}
          G. de Lange$^{1}$, 
          B. Larsson$^{18}$, 
          \changed{D. van Loon$^{1}$,}
          S.C. Madden$^{19}$,
          H. Matsuhara$^{4}$,
          F. Najarro$^{20}$,
          T. Nakagawa$^{4}$, 
          D. Naylor$^{21}$, 
          H. Ogawa$^{4}$,  
          T. Onaka$^{22}$,
          \changed{S. Oyabu$^{13}$,}
          A. Poglitsch$^{23}$, 
          \changed{V. Reveret$^{24}$,} 
          L. Rodriguez$^{24}$, 
          L. Spinoglio$^{11}$,
          \changed{I. Sakon$^{22}$,
          Y. Sato$^{29}$,
          K. Shinozaki$^{29}$,
          R. Shipman$^{1,2}$,
          H. Sugita$^{29}$,
          T. Suzuki$^{13}$,}
          F.F.S. van der Tak$^{1,2}$,
          \changed{J. Torres Redondo$^{20}$,
          T. Wada$^{4}$,
          S.Y. Wang$^{14}$,}
          C.K. Wafelbakker$^{1}$,
          \changed{H. van Weers$^{1}$,
          S. Withington$^{27}$,}
          B. Vandenbussche$^{25}$,
          T. Yamada$^{4}$,
          \changed{I. Yamamura$^{4}$}

\bigskip

Submitted to PASA, 10/10/17

\bigskip
\bigskip
        
}%

\jid{PASA}
\doi{10.1017/pas.\the\year.xxx}
\jyear{\the\year}

\usepackage{aas_macros}
\usepackage{hyperref} 
\hypersetup{colorlinks,citecolor=blue,linkcolor=blue,urlcolor=blue}

\begin{document}

\begin{frontmatter}
\maketitle

\begin{abstract}
Measurements in the infrared wavelength domain allow us to assess directly the physical state and energy balance of cool matter in space, thus enabling the detailed study of the various processes that govern the formation and early evolution of stars and planetary systems in the Milky Way and of galaxies over cosmic time. Previous infrared missions, from IRAS to Herschel, have revealed a great deal about the obscured Universe, but sensitivity has been limited because up to now it has not been possible to fly a telescope that is both large and cold. Such a facility is essential to address key astrophysical questions, especially concerning galaxy evolution and the development of planetary systems. 

SPICA is a mission concept aimed at taking the next step in mid- and far-infrared observational capability by combining a large and cold telescope with instruments employing state-of-the-art ultra-sensitive detectors. The mission concept foresees a 2.5-meter diameter telescope cooled to below 8 K. Rather than using liquid cryogen, a combination of passive cooling and mechanical coolers will be used to cool both the telescope and the instruments. With cooling not dependent on a limited cryogen supply, the mission lifetime can extend significantly beyond the required three years. The combination of low telescope background and instruments with state-of-the-art detectors means that SPICA can provide a huge advance on the capabilities of previous missions.

The SPICA instrument complement offers spectral resolving power ranging from $R\sim$50 through 11000 in the 17-230 $\mu$m domain as well as $R\sim$28.000 spectroscopy between 12 and 18 $\mu$m. Additionally SPICA will be capable of efficient 30-37 $\mu$m broad band mapping, and small field spectroscopic and polarimetric imaging in the 100-350 $\mu$m range. SPICA will enable far infrared spectroscopy with an unprecedented sensitivity of $\sim5\times 10^{-20}$ W/m$^2$ (5$\sigma$/1hr) - at least two orders of magnitude improvement over what has been attained to date. With this exceptional leap in performance, new domains in infrared astronomy will become accessible, allowing us, for example, to unravel definitively galaxy evolution and metal production over cosmic time, to study dust formation and evolution from very early epochs onwards, and to trace the formation history of planetary systems.

\end{abstract}


\end{frontmatter}

%

\section{Introduction}

When identifying strategies for the development of instrumentation for astronomy it is clear that some of the most important themes of current research such as `What are the conditions for planet formation and the emergence of life?' and `How did the Universe originate and what is it made of?' can only be fully answered with observations in the mid- to far-infrared part of the spectrum. This domain is virtually inaccessible from the ground because the Earth's atmosphere is opaque to infrared radiation, and therefore sensitive space-based observations are required which provides the impetus for the SPICA mission concept described in this paper.

 \begin{figure}
	\centering
	\includegraphics[width=1.0\hsize]{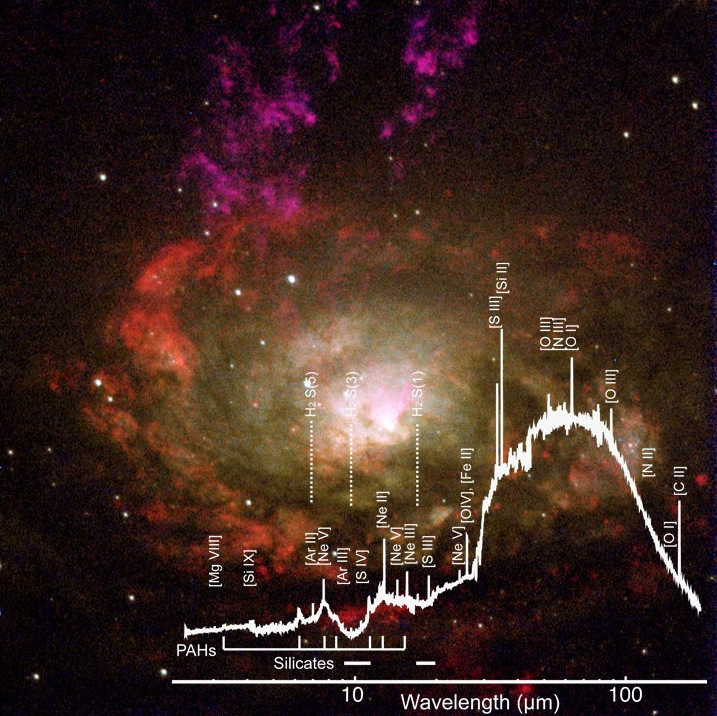}
	\caption{ISO 2-200 $\mu$m spectrum of the Circinus galaxy showing the bright IR peak and the wealth of spectral features including fine structure ionic lines,  molecular lines, and PAH features \citep{1999ESASP.427..825M}. The background shows a Hubble Space Telescope image of the galaxy \citep{Image:Wilson} offering exquisite detail but capturing only a small fraction of the total energy produced by the galaxy - most of which emerges in the mid- and far-IR. 
	}
	\label{Fig:Circinus}
\end{figure}

Over the past three decades, we have come to understand that at least half of the energy ever emitted by stars in galaxies is to be observed in the infrared \citep[see e.g.][]{2006A&A...451..417D}. Observations at IR wavelengths are optimal for the study of galactic evolution in which peak activity occurs at redshifts of z$\sim$1-4, when the Universe was roughly 2-3 Gyr old - as was concluded primarily through deep and wide-field observations with previous infrared space observatories: IRAS \citep{1984ApJ...278L...1N}, ISO  \citep{1996A&A...315L..27K}, {\it Spitzer} \citep{2004ApJS..154....1W}, AKARI \citep{2007PASJ...59S.369M}, {\it Herschel} \citep{2010A&A...518L...1P} and WISE  \citep{2010AJ....140.1868W}. In addition, many of the basic processes of star formation and evolution, from pre-stellar cores to the clearing of gaseous proto-planetary discs, and the presence of dust excess around main sequence stars, were discovered by these pioneering missions. Notwithstanding the success of these  missions, these observatories either had small cold telescopes, or large, warm mirrors, ultimately limiting their ability to probe the physics of the faintest and most distant, obscured sources that dominate the mid- and far- infrared emission in our Universe. 

\begin{figure}
	\centering
	\includegraphics[width=1.0\hsize]{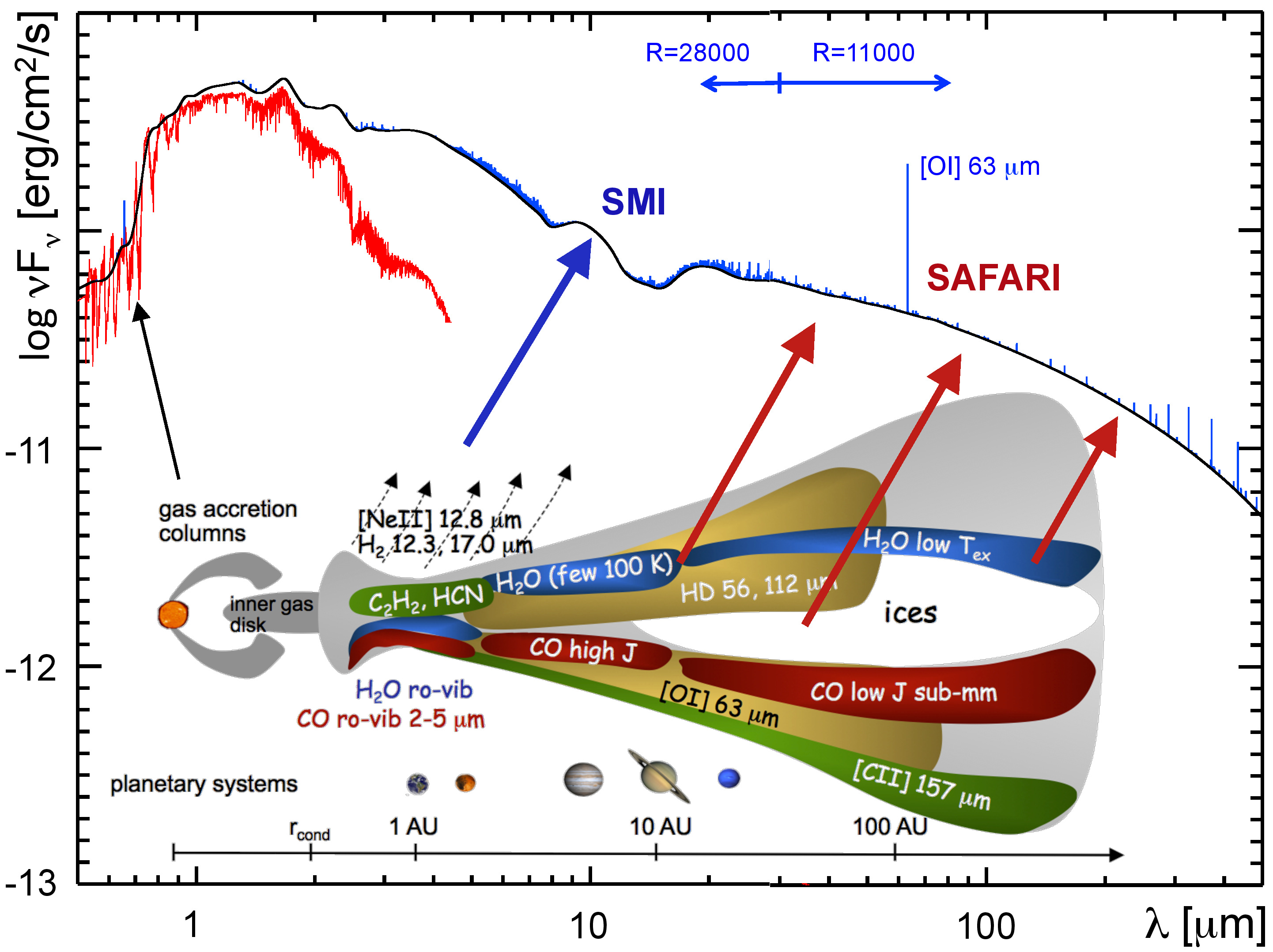}
	\caption{Model SED for a protoplanetary disc, illustrating that the bulk of the planet-forming reservoir is best studied at mid- to far-IR wavelengths, but requires high sensitivity for the detection of gas lines and dust/ice features.}
	\label{Fig:PPD-SED}
\end{figure}

\subsection{The power of infrared diagnostics}

\begin{figure}
	\centering
	\includegraphics[width=1.0\hsize]{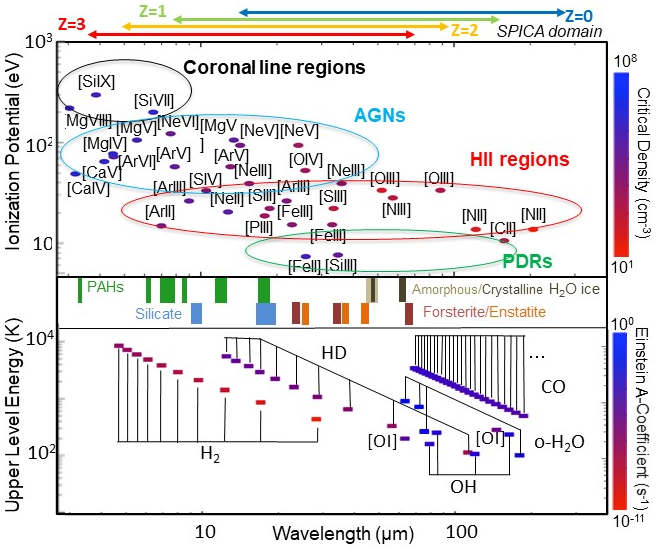}
	\caption{Upper panel: ionization potential versus wavelength for key infrared ionic fine-structure lines \citep{1992ApJ...399..504S}. Lower panel: upper energy level of molecular transition and spectral features from PAHs, water ice bands, and other species versus wavelength. The SPICA domain is indicated above for several different redshifts.}
	\label{Fig:IRTransitions}
\end{figure}

The mid- to far-IR spectral range hosts a suite of ionic, atomic, molecular and dust features covering a wide range of excitation, density and metallicity, directly tracing the physical conditions both in the nuclei of galaxies and in the regions where stars and planets form (see Figures~\ref{Fig:Circinus}, \ref{Fig:PPD-SED} and \ref{Fig:IRTransitions}). Ionic fine structure lines (e.g. [NeII], [SIII], [OIII]) probe HII regions around hot young stars, providing a measure of the star formation rate, stellar type, and the density of the gas. Lines from highly ionized species (e.g. [OIV], [NeV]) trace the presence of energetic photons emitted from AGN, providing direct measures of the accretion rate. Photo-dissociation regions (PDR), the transition between young stars and their parent molecular clouds, can be studied via the strong [CII] and [OI] lines and the emission from small dust grains and PAHs. The major coolants of the diffuse warm gas (e.g. [NII]) in galaxies also occur in the  far-infrared giving us a complete picture of the ISM.

The rest-frame infrared furthermore is home to pure rotational H$_2$, HD and OH lines (including their ground state lines) and mid- to high-J CO and H$_2$O lines, most notably the H$_2$O ground state line at 179 $\mu$m. Finally, the strong PAH emission features (carrying 1-10\% of the total IR emission in star-forming galaxies) with their unique spectral signature, can be used to determine redshifts of galaxies too dust-obscured to be detected at shorter wavelengths. The infrared also hosts numerous unique dust features from minerals such as olivine, calcite and dolomite that probe evolution from pristine to processed dust (e.g. aqueous alteration), as well as CO$_2$ ice and molecules like C$_2$H$_2$ and fullerenes. A table listing these various spectral tracers can be found in  \citet{WP7-vanderTak}.

\begin{figure}
	\centering
	\includegraphics[width=0.9\hsize]{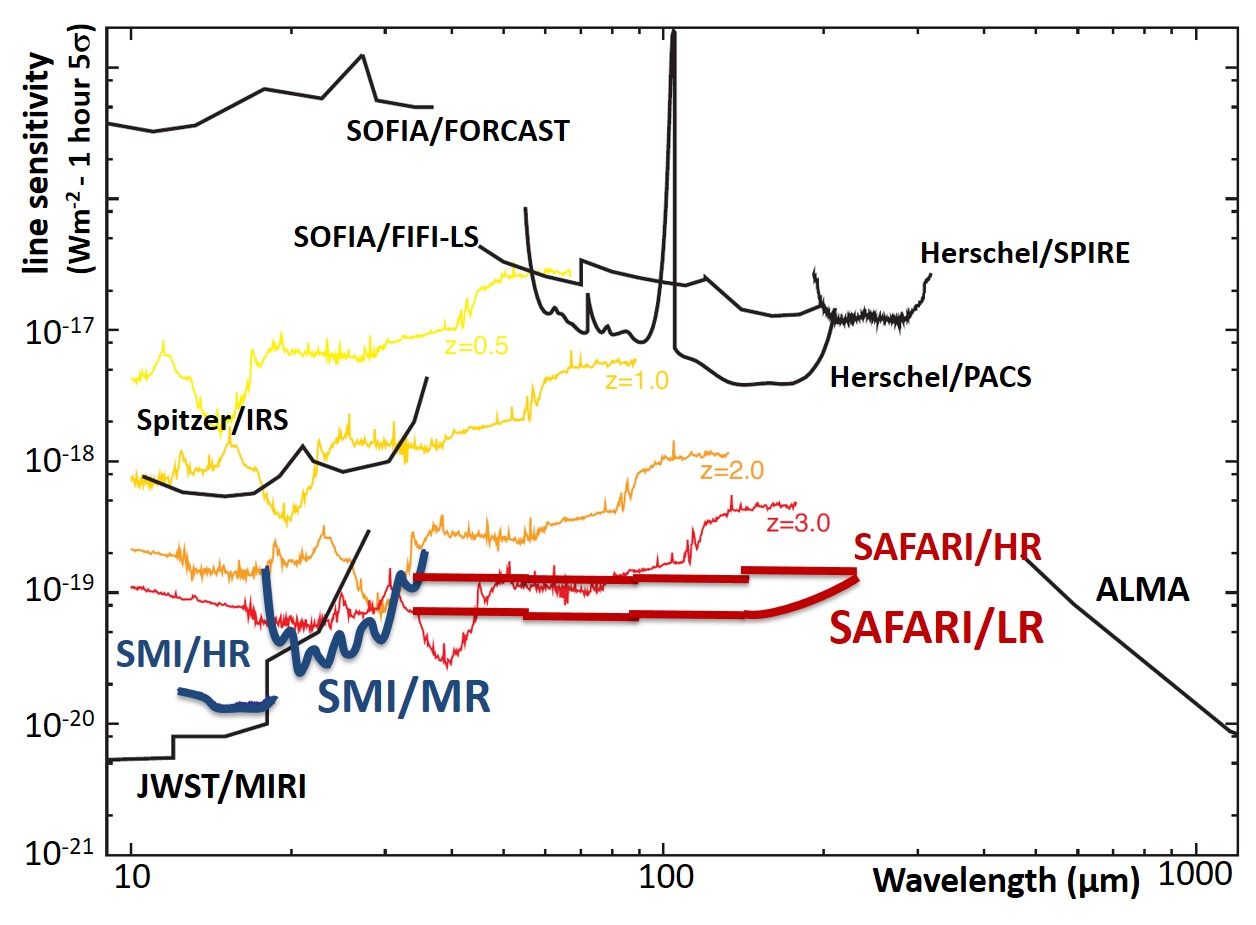}
	\caption{Projected spectroscopic sensitivity of the SPICA instruments as compared to other infrared facilities \changed{(at the SAFARI/LR resolution of $\sim$300)}. The SAFARI sensitivity assumes a detector NEP of $2\times10^{-19}$W/$\surd$Hz. The infrared spectrum of the Circinus galaxy, scaled to L=$10^{12}$ L$_\odot$, for redshifts 0.5 to 3, and smoothed to the SAFARI/LR resolution, is superimposed.}
	\label{Fig:SPICASensitivities}
\end{figure}

Taking advantage of  progress in detector performance and cryogenic cooling technologies, we now stand on the threshold of unprecedented advances in our ability to study the hidden, dusty Universe. An observatory like SPICA as presented in this paper, with  a large, cold telescope complemented by an instrument suite that exploits the sensitivity attainable with the low thermal background, can in the mid/far-infrared achieve a gain of {\it over two orders of magnitude} in spectroscopic sensitivity as compared to Herschel, Spitzer and SOFIA \citep[][see also Figure~\ref{Fig:SPICASensitivities}]{2016SPIE.9973E..0IB}. In addition such a mission will provide access to wavelengths well beyond those accessible with the James Webb Space Telescope \citep[JWST][]{2006SSRv..123..485G} and the new generation of extremely large telescopess \citep{2016SPIE.9906E..0WT,2016SPIE.9906E..12M,2016APS..APR.E7002C}. Sitting squarely between JWST  and ALMA \citep{2009IEEEP..97.1463W}, SPICA will enable the discovery and detailed study of the coldest bodies in our Solar System, emerging planetary systems in the Galaxy, and the earliest star forming galaxies and growing super-massive black holes - a gigantic leap in capabilities for unveiling the `hidden Universe'.  With its large cryogenic infrared space telescope, SPICA will allow astronomers to peer into the dust-enshrouded phases of galactic, stellar and planetary formation and evolution, revealing the physical, dynamical and chemical state of the gas and dust, and providing answers to a range of fundamental astronomical questions:

\begin{itemize}

\item What are the roles of star formation, accretion onto and feedback from central black holes and supernova explosions in shaping galaxy evolution over cosmic time?

\item How are metals and dust produced and destroyed in galaxies? How does the matter cycle within galaxies and between galactic discs, halos and intergalactic medium?

\item How did primordial gas clouds collapse into the first galaxies and black holes?

\item What is the role of magnetic fields at the onset of star formation in the Milky Way?

\item When and how does gas evolve from primordial discs into emerging planetary systems?

\item How do ices and minerals evolve in the planet formation era, as seed for Solar Systems, acting as the seeds for planet formation?

\end{itemize}

These questions, their background science, and how SPICA would help resolve them, are discussed in much more detail in a dedicated series of papers \citep{WP1-Spinoglio,WP2-Fernandez-Onteveros,WP3-Gonzales-Alfonso,WP4-Gruppioni,WP5-Kaneda,WP6-Egami,WP7-vanderTak,WP8-Andre,2017A&A...605A..69T,2017ApJ...836..118N,2016ApJ...827..113N,KampInprep}.

\begin{figure}
	\centering
	\includegraphics[width=0.8\hsize]{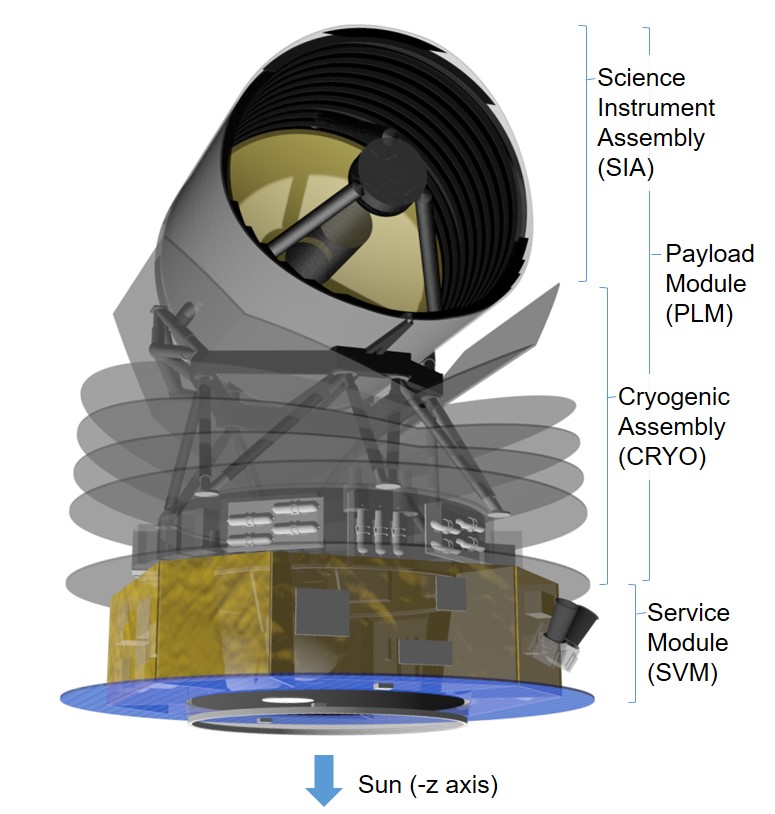}
	\caption{The SPICA spacecraft configuration. The scientific instruments are mounted on the optical bench on the rear of the telescope (see also Figure~\ref{Fig:OpticalBench}).
	}
	\label{Fig:Spacecraft}
\end{figure}

\subsection{A cryogenic infrared space telescope}

Early mission concepts for a cryogenic infrared space telescope, initially proposed by the Japanese space agency JAXA, have been described extensively elsewhere \citep{1998SPIE.3356..462N, 2009ExA....23..193S, 2014SPIE.9143E..1IN}. Over time the  mission concept has evolved significantly, SPICA is now envisaged as a joint European-Japanese project to be implemented for launch and operations at the end of the next decade. \changed{As an ESA mission the project would go to through the standard ESA phasing and milestones, starting with the detailed Phase A study in 2019/2020 and mission selection in early 2021.} A joint ESA-JAXA study was already conducted  to assess the technical and programmatic feasibility of possible mission configurations \citep{Report:CDFCryoTel}, the satellite configuration has since been further optimized. 

In the currently forseen architecture, building on heritage from the ESA Planck mission \citep{2011A&A...536A...1P}, the optical axis of the  2.5 meter diameter cold telescope is perpendicular to the axis of the spacecraft and `V-groove' radiators, mounted between the telescope assembly and the satellite service module, provide very efficient passive cooling. The solar cells to provide power are mounted on the bottom side of the service module, which is always orientated towards the Sun. In parallel with this optimisation of the mission concept, the science payload has also been revisited and upgraded. The resulting mission concept as considered in this paper will provide extremely sensitive - of order 5$ \times 10^{-20}$ W/m$^2$ (5$\sigma$/1hr) - spectroscopic capabilities from 17 through 230 $\mu$m at resolutions ranging from $R=50$ through 11000. A high resolution $R\sim$28.000 capability is provided for the 12-18 $\mu$m wavelength range. Efficient broad band photometric mapping can be carried out  in the 30-37 $\mu$m domain, as can spectroscopic imaging for small fields in the 35-230 $\mu$m range. In addition a polarimetric imaging capability is provided in three  bands at 100, 200 and 350 $\mu$m. The spectroscopic sensitivity (Figure~\ref{Fig:SPICASensitivities}) of the instrument suite will typically provide two orders of magnitude improvement over what has been attained to date, corresponding to a truly enormous increase in observing speed.  Such an exceptional leap in performance is bound to produce many scientific advances. Some of these are predictable today and form the core science case for a SPICA concept and will drive its design. However, many additional advances are impossible to predict, and the discovery space is undoubtedly large in such a mission.

\begin{figure}
	\centering
	\begin{minipage}{0.49\hsize}
		\includegraphics[width=\linewidth]{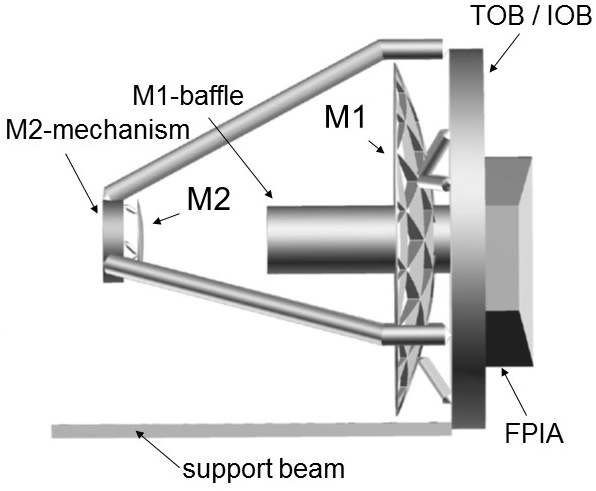}
	\end{minipage}
	\begin{minipage}{0.49\hsize}
		\includegraphics[width=\linewidth]{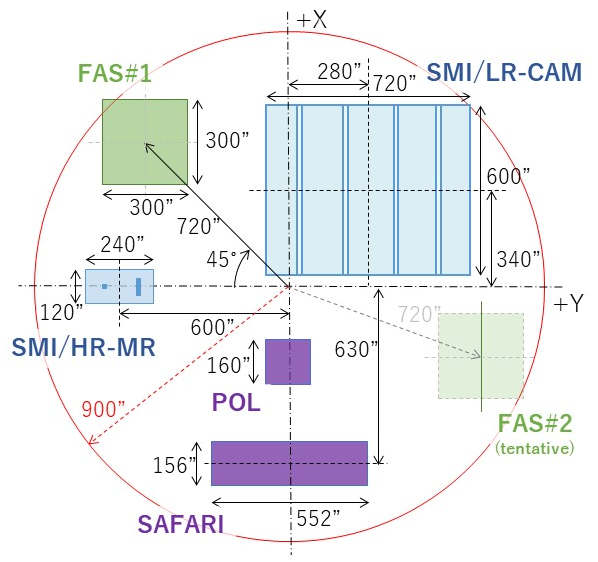}
	\end{minipage}
	\caption{Left: configuration of the SPICA telescope assembly. The scientific instruments are mounted in Focal Plane Instrument Assembly (FPIA), on the optical bench on the rear of the telescope. Right: the SPICA instrument focal plane layout.}
	\label{Fig:OpticalBench}
\end{figure}

\section{The SPICA satellite concept}

The SPICA mission concept utilises a 2.5-meter class Ritchey-Chr\'{e}tien telescope, cooled to below 8 K. The telescope is mounted on the service module with its axis perpendicular to the spacecraft axis. The practical telescope size is primarily limited by the launcher capabilities - fairing diameter and mass capability. The telescope and instrument suite, which are never exposed to direct sunlight, will be cooled using mechanical coolers in combination with V-groove radiators. Solar panels to provide electrical power are mounted on the bottom of the service module, which is directed towards the Sun.  The nominal mission lifetime is three years, with a goal of five years, but as cooling is provided by mechanical coolers the operational lifetime in practice would not be limited by liquid cyrogen consumables but only by the available propellant and the durability of mechanisms and on-board electronics. 

The overall configuration of the  spacecraft concept is shown in Figure~\ref{Fig:Spacecraft}, with the service module (SVM) housing the general spacecraft support functions below, and on the top the payload module (PLM) with the Science Instrument Assembly (SIA), and the Cryogenic Assembly (CRYO) housing the passive and active cooling system for the SIA. Figure~\ref{Fig:OpticalBench} shows the telescope configuration on the left and on the right a tentative focal plane aperture assignment for the instruments and the two Focal Plane Attitude Sensors (part of the overall spacecraft pointing system). The main parameters of the satellite are summarised in Table~\ref{tab:MissionParameters}.

\begin{table}
	\caption{Main SPICA mission parameters}
\begin{tabular}{p{0.02\hsize} p{0.88\hsize}}
	\hline\hline 
	\multicolumn{2}{l}{Item  Specification}  \\ 
	\hline
	\multicolumn{2}{l}{\it Spacecraft system}\\ 
		& Height: $\sim$ 5.9 meter \\
		& Diameter: $\sim$ 4.5 meter \\
		& Mass including consumables: 3.65 tonnes \\
                     & Launcher: JAXA H3 \\
		& Attitude control: 3-axis stabilized with startracker, gyro and fine attitude sensors \\ 
		& Absolute Pointing Error $\sim$ 0.5" \\
		& Power: $\sim$ 14 m$^2$ solar array providing 3 kW \\
		& Data handling: 24 hour autonomous operation, 100 GB on-board data storage, X-band downlink at $\sim$ 10 Mbps \\  
	\multicolumn{2}{l}{\it Cooling system} \\
		& Passive cooling combined with mechanical coolers \\
		& End of life cooling power: \\
		& \hspace{6pt} Stirling coolers: > 200 mW  at 20K  \\
		& \hspace{6pt} 4K Joule-Thomson coolers:  40 mW   at 4.5K \\
		& \hspace{6pt} 1K Joule-Thomson coolers:  10 mW   at 1.7K \\
          \multicolumn{2}{l}{\it Telescope}\\ 
		& 2.5 meter Ritchey-Cr\'{e}tien \\
		& Strehl ratio for telescope/instruments > 0.80 at 20 $\mu$m \\ 
		& Cooled below 8 K \\ 
	\multicolumn{2}{l}{\it Instruments}\\ 
		& Mid-infrared spectroscopy 12-36 $\mu$m - SMI \\ 
		& Far-infrared spectroscopy - 34-250 $\mu$m - SAFARI \\
                     & Mid-infrared imaging 30-37 $\mu$m - SMI \\ 
		& Far-infrared imaging polarimetry 100/200/350 $\mu$m - POL \\
	\hline\hline 
\end{tabular} 
    \label{tab:MissionParameters}
\end{table}

The SPICA telescope derives directly from the Herschel mission, in design as well as implementation. The system optical design is based primarily on the ESA CDF study \citep{Report:CDFCryoTel} with further follow up work at JAXA. The overall layout for the telescope optics is shown in Figure~\ref{Fig:OpticalBench}. The telescope primary and secondary mirrors and the secondary supports will be made of silicon carbide (SiC), exploiting the heritage of the Herschel, AKARI and Gaia missions in terms of structure and technology and materials. The entire telescope assembly will be cooled down to < 8 K.  Because of the extremely low power radiant levels at the detectors, stray-light  rejection will be an important consideration in the detailed design of the telescope baffle and the instrument optics. Estimates for the wave-front error budget indicate that the secondary assembly will require a three-axis (focusing, tip, and tilt) correction mechanism, which can be driven by actuators similar to those used on the Gaia spacecraft \citep{2016A&A...595A...1G}.

\subsection{The cryogenic assembly}

The Payload Module is connected by trusses to the cryogenic assembly, consisting of the Cooler Module and the Thermal Insulation and Radiative Cooling System. The Cooler Module houses the mechanical cryo-cooler units with their electronics, and the warm electronics modules of the science instruments. The cryogenic assembly is designed with the primary goal to cool the telescope assembly (STA) to below 8 K, and to provide cold temperature stages  for the instruments at 4.8 K and 1.8 K. The system is designed to reach a steady state for those temperatures within 180 days after launch. The SPICA cooling system combines radiative cooling with V-Groove shields similar to the Planck design \citep{2011A&A...536A...1P} with mechanical cryocoolers \citep{2010Cryo...50..566S,2014Cryo...64..228S}. In addition, a telescope shield, which is actively cooled down to 25 K, is placed between the V-Grooves and the telescope baffle \citep{2016SPIE.9904E..2HO}.

\begin{figure}
	\centering
	\includegraphics[width=0.9\hsize]{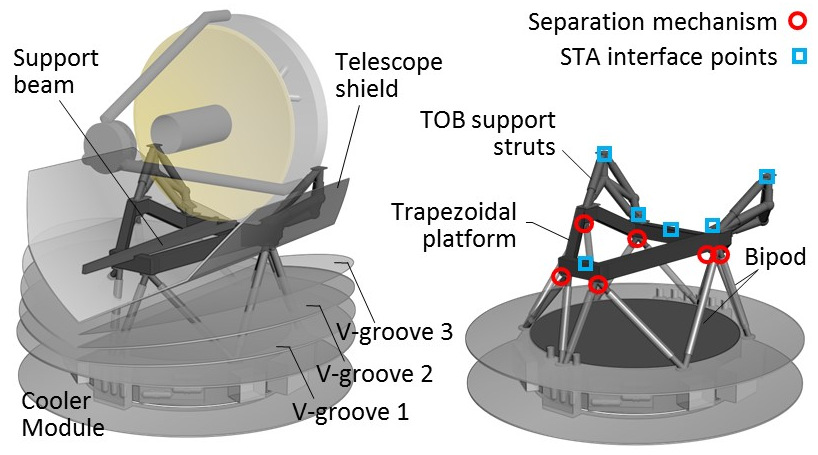}
	\caption{External view of the SPICA cryogenic support structure with bipods, V-groove shields and trapezoidal mounting platform for the telescope assembly.
	}
	\label{Fig:CryoStructure}
\end{figure}

The spacecraft will keep its attitude always with the -Z axis directed towards the Sun (see Figure~\ref{Fig:Spacecraft}), such that only the solar panel on the bottom of the SVM is illuminated and the SIA is never exposed to direct solar radiation. Only the actively cooled telescope  shield and deep space are within the SIA field of view.

\begin{figure*}
	\centering
	\begin{minipage}{0.45\hsize}
		\centering
		\includegraphics[width=\linewidth]{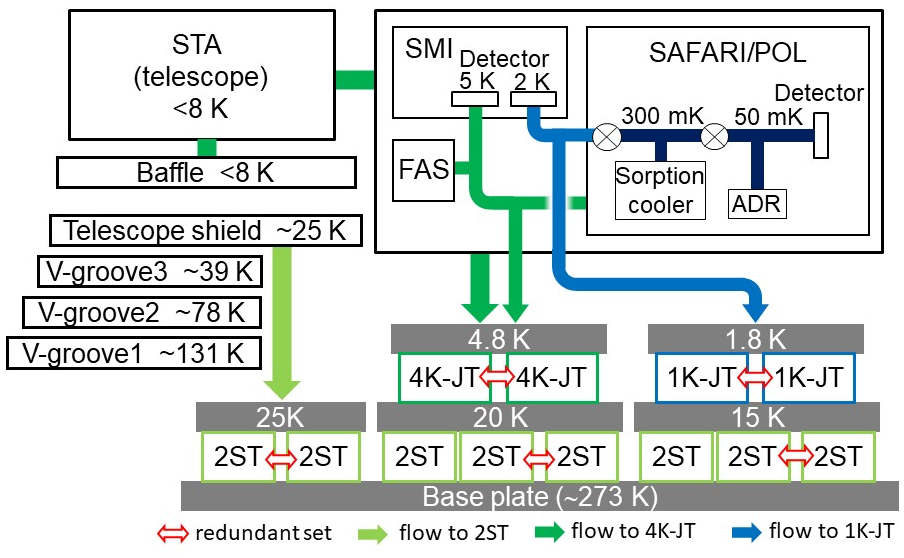}
	\end{minipage}
    \hspace{0.05\hsize}
	\begin{minipage}{0.45\hsize}
		\centering
		\includegraphics[width=\linewidth]{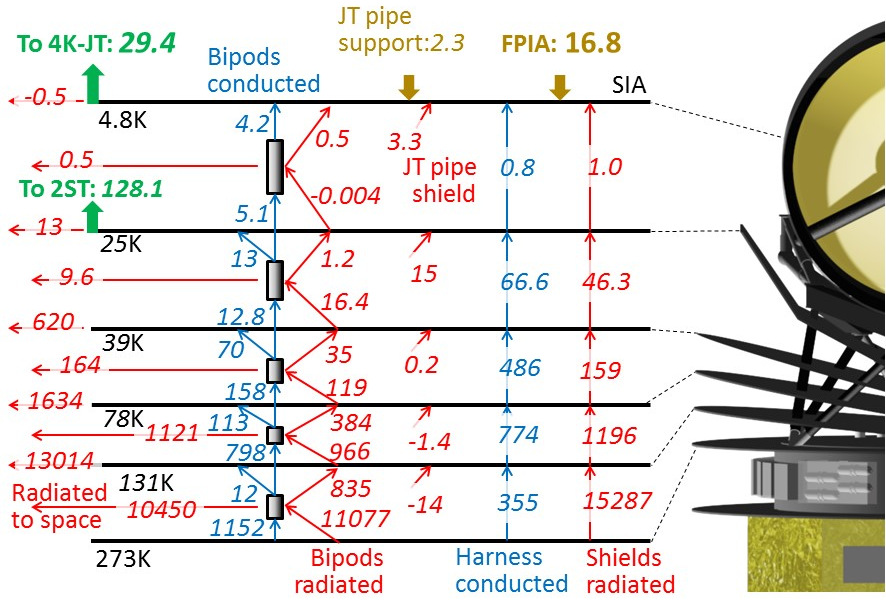}
	\end{minipage}
	\caption{The SPICA cryogenic system. Left: the cooling chain concept. 4K-JT and 1K-JT coolers are used to cool the STA and instrument focal plane units, and used as precoolers for the SAFARI and POL sub-K coolers (Sorption cooler and adiabatic demagnetization refrigerator - ADR). The redundant chains provide a safeguard against failures in any one of the coolers. Right: a heat flow diagram (values in mW). With the maximum heat load (16.8 mW) from the instruments, the estimated heat flow to the 4K-JT is 29.4 mW, well below the 40 mW end-of-life cooling capability of the 4K-JT system with one failed cooler}
	\label{Fig:SPICACryo}
\end{figure*}

The proposed cryogenic assembly is shown in Figure~\ref{Fig:CryoStructure}. A trapezoidal platform made of carbon fibre reinforced plastic (CFRP), with six CFRP bipods, forms a stable, light-weight structure acting as a mounting platform for the cold telescope. The struts of the telescope optical bench support structure are made of low-thermal-conductivity CFRP to reduce the conducted heat load from the warm SVM. The bipods and telescope optical bench support need to be substantial enough to satisfy the spacecraft stiffness requirements and withstand the launch loads. However, they  will also be the main conductive heat path from SVM to SIA, requiring very low thermal conductance.  Therfore an on-orbit truss separation mechanism \citep{2015JATIS...1b7001M} will  be used  to de-couple strong supports after  launch, when high mechanical strength will no longer be necessary, leaving only the low-conductance supports. Six interface points between bipods and the platform (indicated with red circles in Figure~\ref{Fig:CryoStructure}) will be separated using a CFRP octagonal spring mechanism leaving the SVM and SIA connected only by the low-thermal-conductance elements. Each of the three V-groove shields, attached to the bipods, is constructed of sandwich panels with aluminium face sheets and an aluminium honeycomb core. The telescope shield is an aluminium shell or full aluminium sandwich panel, supported by the bipods.

The SPICA cooling chain concept \citep{2016SPIE.9904E..3WS} is shown in Figure~\ref{Fig:SPICACryo}. The system has two chains for the SIA, one to provide a 4.8-K level and the other to provide a 1.8-K level.  Both chains use two types of Joule Thomson coolers (4K-JT and 1K-JT) in combination with three double-stage Stirling (2ST) pre-coolers. The telescope shield is cooled to about 25 K by two dedicated 2ST coolers, and three V-groove shields are used to reduce the  heat load to the telescope shield. Redundant chains are provided as safeguard against failures in any one of the coolers. 

One of the challenges for the SPICA project will be the validation of the SPICA cryogenic system, as its performance is of prime importance for the success of the mission. For this validation an extensive ground test program will need to be executed in which the full payload - i.e. the integerated telescope {\it and} instrument assemblies - as well as relevant test sources are operated under flight-like conditions.  

 \begin{figure}[b]
	\centering
	\includegraphics[width=0.6\hsize]{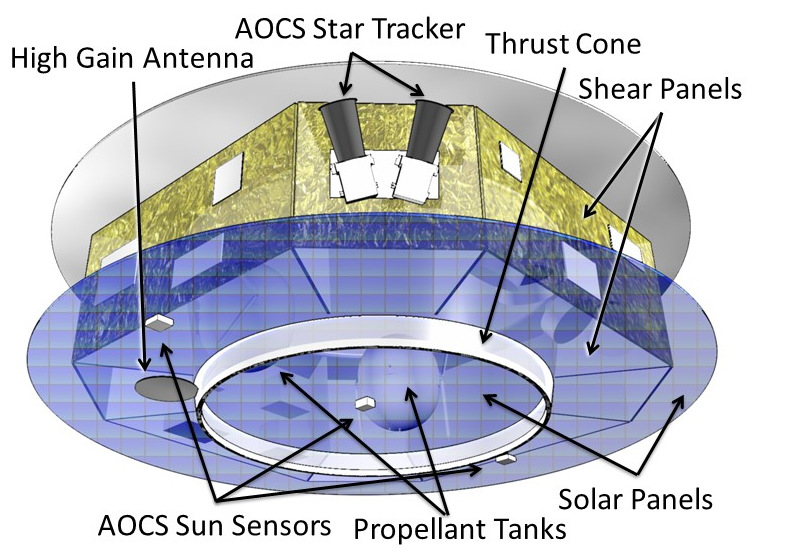}
	\caption{The SPICA service module (SVM).
	}
	\label{Fig:SVM}
\end{figure}

\subsection{Service Module (SVM)}

The SPICA SVM (see Figure~\ref{Fig:SVM}), housing all spacecraft support functions, is considered fairly standard and can be derived directly from e.g. the Herschel and Planck missions. It will be a combination of a thrust cone as load path between PLM and the launcher, and shear panels. The top and bottom platforms and the side panels accommodate spacecraft equipment. The solar panels are mounted on the lower platform and always face the  Sun. To maintain the units inside the SVM at their nominal operating temperature, the bottom and top platforms are insulated from the rest of SVM by means of Multi Layer Insulation (MLI) and thermal stand-offs. The top surface facing the PLM is also covered with MLI. 

The baseline data handling system architecture comprises of an On Board Compute, Remote Terminal Units and Solid State Mass Memory. The platform generates about 15 kbps housekeeping data rate, and the science instruments generate data at an average rate of 3 Mbps with around 15 kbps of instrument housekeeping. The mass memory will need to be able to store up to 72 hours of spacecraft data (rougly 100 GB). To a large extent the SVM sub-systems, also including the Command and Data Handling subsystem, are based on highly recurrent designs and little need for technology developments is expected. Nominal telemetry and telecommand operation as well as scientific telemetry downlink use X band frequencies. Low gain antennas are positioned such that continuous coverage is achievable for low data rate command and housekeeping. A high gain antenna is mounted on the bottom panel of the spacecraft which in nominal attitude points towards the Earth. The proposed architecture will have active redundancy for the uplink and passive redundancy for the high data rate downlink.

The Attitude and Orbit Control Subsystem (AOCS) will use a three-axis stabilized system based on a star-tracker and gyro estimation filter for coarse pointing modes, with a design derived directly from the Herschel AOCS architecture. In addition, a Focal-plane Attitude Sensors (FAS) is likely to be needed  to fulfil the more stringent requirements in some of the science fine pointing modes. Two (redundant) FAS cold units, housing the camera optics and 1K$\times$1K near-IR detectors, will be mounted on the SIA, with warm signal processing electronics in the SVM. The FAS has a 5'$\times$5' FoV (see Figure~\ref{Fig:Spacecraft}), which can track at least five stars at any one time. For science observations the AOCS will be required to achieve a pointing accuracy of order $0.5-1$", depending on the details of the observing mode. The absolute pointing error budget contributions come from three different main sources: the structure misalignment between the instruments and FAS, the FAS constant bias, and controller performance including actuator disturbance noise. The relative pointing error is mainly due to three contributors: attitude relative pointing estimates (FAS + gyros), short term controller performance including actuator noise, and $\mu$-vibration sources (reaction wheels and cryocoolers). The a-posteriori absolute pointing knowledge error budget is linked to the attitude estimate error achievable on board with possible corrections obtained by post-processing on ground.

\section{Nominal instrument complement}

\begin{figure}[t]
	\centering
	\includegraphics[width=\hsize]{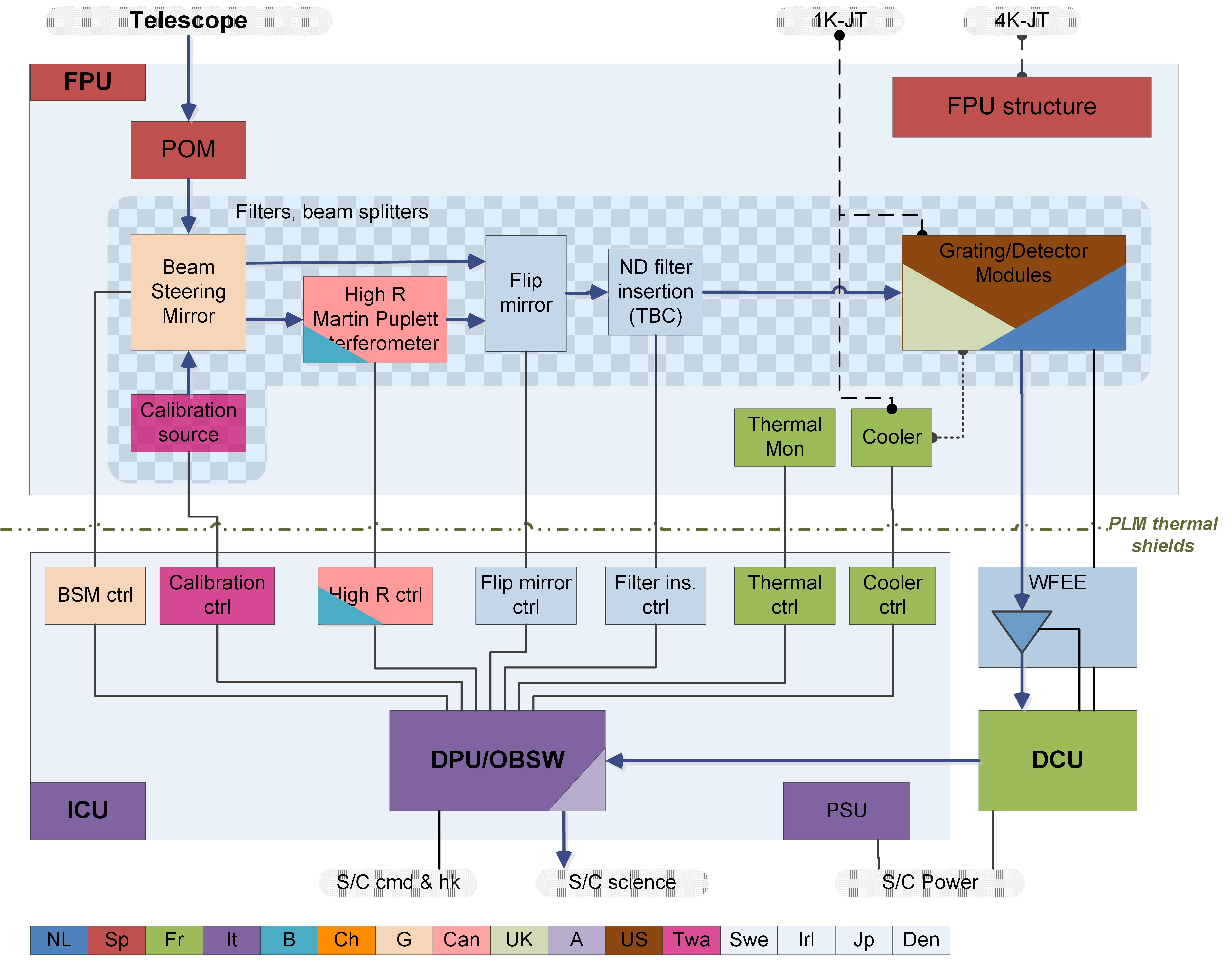}
	\caption{The functional block diagram for SAFARI. The top part represents the Focal Plane Unit mounted on the 4.8 K instrument optical bench. The bottom part shows the warm electronics, mounted on the CRYO assembly. }
	\label{Fig:SAFARIBlockdiagram}
\end{figure}

\subsection{A far infrared spectrometer}

With galaxy evolution over cosmic time as a main science driver, the SPICA far infrared spectrometer SAFARI is  optimised primarily to achieve the best possible sensitivity, within the bounds of the available resources (thermal, number of detectors, power, mass), at a moderate resolution of $R\sim$300, with instantaneous coverage over the full 34 to 230 $\mu$m range. A secondary driver is the requirement to  study line profiles at higher spectral resolution, e.g. to discern the in-fall and outflow of matter from active galactic nuclei and star-forming galaxies. This leads to the implementation of an additional high resolution mode using a Martin-Puplett interferometer \citep{MartinPuplett}. With this design, the sensitivity of the $R\sim$300 SAFARI/LR mode will be about $5 \times 10^{-20}$ W/m$^2$ (5$\sigma$, 1hr) assuming a TES (Transition Edge Sensor) detector NEP (Noise Equivalent Power) of $2 \times 10^{-19}$ W/$\surd$Hz. With such high sensitivity astronomers will be able to characterise, over the full spectral band, average-luminosity galaxies ($\sim$L$_*$)  out to redshifts of at least 3. It should be noted that in this design concept, with essentially zero background emission from the telescope and instruments, further advances in detector sensitivity translate directly to better overall instrument performance.

\begin{table}
	\caption{SAFARI performance summary.}
	\setlength{\tabcolsep}{1pt}
	\begin{tabular}{lcccc}
		\hline \hline
	Band	& SW & MW  & LW  & VLW \rule{0pt}{11pt} \\ 
		\hline 
	$\lambda$ range	& 34-56$\mu$m & 54-89$\mu$m & 87-143$\mu$m & 140-230$\mu$m \rule{0pt}{10pt} \\ 
	high $R$  & 11700-7150 & 7400-4500 & 4600-2800 & 2850-1740 \\
	nom. $R$  & 300 & 300 & 300 & 300 \\
	FWHM	& 4.5" & 7.2" & 12" & 19"  \\ 
		\hline 
	\multicolumn{5}{l}{ Point source spectr. 5$\sigma$-1hr flux limit ($10^{-20}$ Wm$^{-2}$) \rule{0pt}{10pt} } \\ 
	high $R$	& 13 & 13  & 13 & 15 \\ 
	nom. $R$ 	& 7.2  & 6.6  & 6.6 & 8.2  \\  
	\multicolumn{5}{l}{Mapping spectr. 1'$\times$1' 5$\sigma$-1hr flux limit ($10^{-20}$ Wm$^{-2}$)} \\ 
	high $R$	& 189 & 113  & 73 & 51 \\ 
	nom. $R$	& 84  & 49  & 30 & 23  \\ 
		\hline 
	\multicolumn{5}{l}{Mapping phot. $1'\times1'$ 5$\sigma$-1hr flux density limit ($\mu$Jy)\rule{0pt}{10pt} } \\
						& 209 & 192 & 194 & 239 \\
	5$\sigma$ conf.		& 15  & 200 & 2000 & 10000 \\
		\hline\hline
          \multicolumn{5}{l}{high $R$ - high resolving power mode; $\mu$m \rule{0pt}{11pt}} \\
          \multicolumn{5}{r}{ $R\sim 11000$ at 34  to $R\sim1500$ at 230 $\mu$m } \\
          \multicolumn{5}{l}{nom. $R$ - nominal resolving power; $R\sim300$ } \\
          \multicolumn{5}{l}{5$\sigma$ conf. - 5$\sigma$ confonfusion limit } \\
	\end{tabular} 

	\label{tab:SAFARI}
\end{table}

A functional block diagram for the SAFARI spectrometer system is shown in Figure~\ref{Fig:SAFARIBlockdiagram}. It illustrates the division of the system into two major elements on the spacecraft; the top of the diagram shows the cold Focal Plane Unit (FPU) mounted on the optical bench (at 4.8 K) and the lower part the warm electronics mounted on the CRYO assembly (see also Figure~\ref{Fig:Spacecraft}). The signal from the telescope is fed into the instrument by a pick-off mirror in the common instrument focal plane (see Figure~\ref{Fig:OpticalBench}). The baseline SAFARI design uses a 2D beam steering mirror (BSM) in an Offner relay to send the incoming signal to the dispersion and detection optics. The BSM can be used to select either the sky or an internal calibration signal, and convey this to either the nominal $R\sim$300 (SAFARI/LR) resolution optics chain or to the $R\sim$2000-11000 resolution SAFARI/HR optics chain. The full 34-230 $\mu$m wavelength range is covered by four separate bands, each with its own grating and detector array. The signal is conditioned and split over these four wavelength bands, using band defining filters and dichroics, for input into the Grating Modules (GM, see sec. \ref{sec:GM}) which house the grating optics and the band detector modules. The spectrometer specifications and capabilities are listed in Table~\ref{tab:SAFARI}.

The operation of SAFARI will be controlled by an Instrument Control Unit (ICU). The ICU will receive instrument commands from the spacecraft, interpret and validate them, and subsequently operate all the different SAFARI units in accordance with these commands. In parallel the ICU will collect housekeeping data from the various subsystems and use these to monitor the instrument health and observation progress. The ICU will also collect the science data preconditioned in the Warm Front End Electronics (WFEE) and acquired by the Detector Control Unit (DCU), and package these data along with the housekeeping information for transmission to the spacecraft mass memory from where it will be subsequently downloaded to the ground.

\begin{figure}
	\centering

	\begin{minipage}{0.6\hsize}
		\centering
		\includegraphics[width=\hsize]{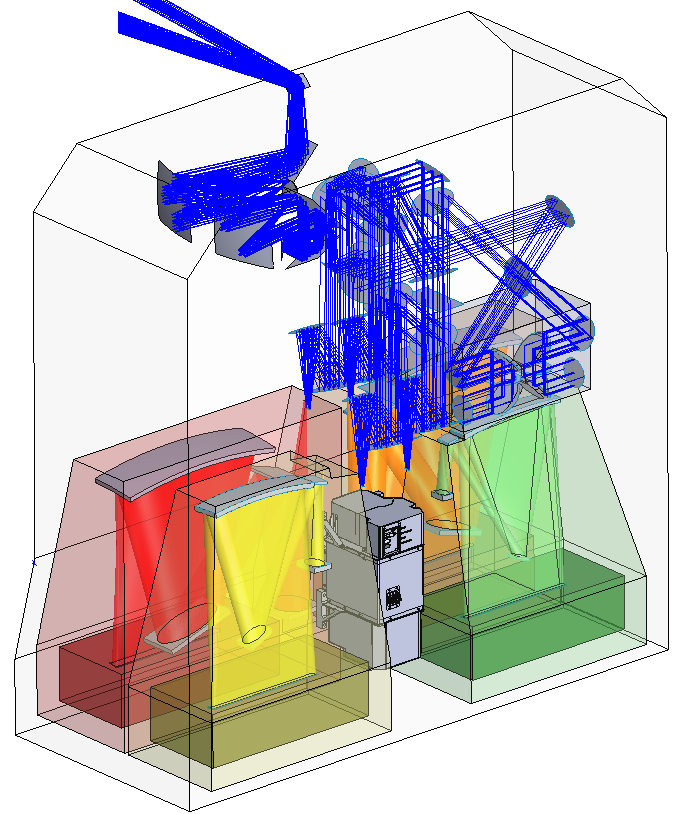}
	\end{minipage}
	\begin{minipage}{0.35\hsize}
	\centering
	\includegraphics[width=\hsize]{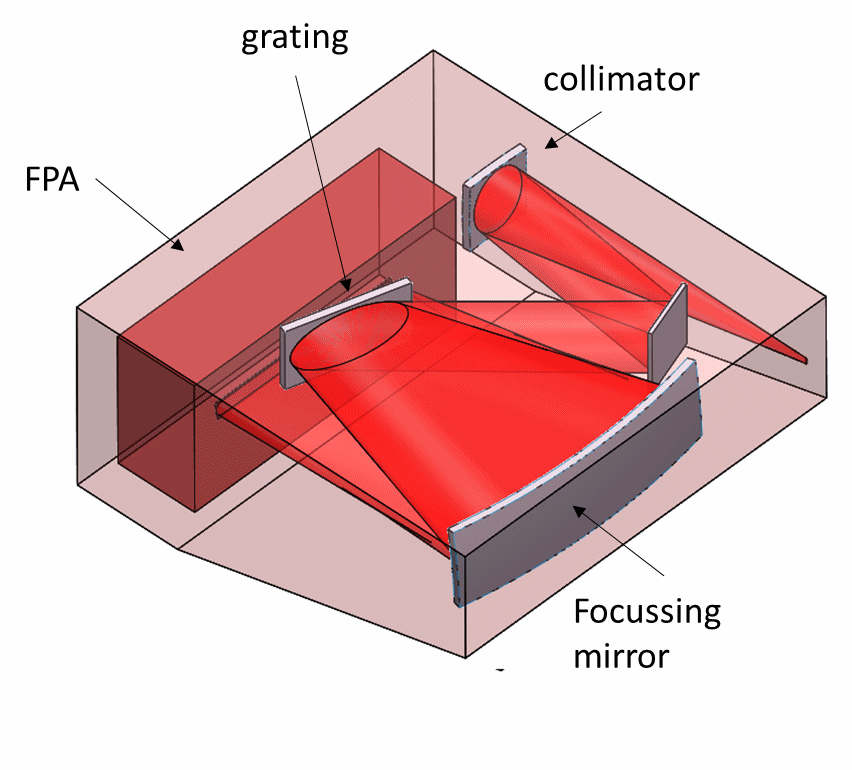}
	\includegraphics[angle=90,width=0.8\hsize]{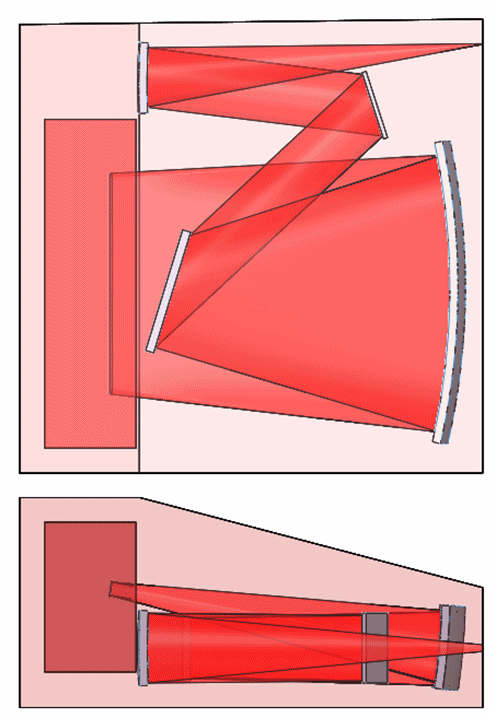}
	\end{minipage}
	\caption{Left: the SAFARI Focal Plane Unit (FPU) as it is mounted against the back of the telescope. The beam from the telescope secondary comes from the top left and is sent into the instrument via the pick-off mirror on the top of the instrument box. From there it goes into the Offner relay optics and on to the Beam and Mode Distribution Optics. On the right the Martin-Puplett signal path and its moving mirror stage can be seen. Three of the four grating modules (red: VLW, yellow: MW and green: SW) are visible on the bottom, the LW band GM (orange) is at the back. Between the MW and SW grating modules the cooler unit (grey) is visible. Right: schematic views of the SAFARI Very Long Wavelength band Grating Module. The location of the detector module as shown in Figure~\ref{Fig:SAFARIFPA} is indicated by the rectangular box denoted FPA.
	}
	\label{Fig:SAFARIFPU}
\end{figure}

\subsubsection{The SAFARI detector grating modules}
\label{sec:GM}

The low resolution dispersion is effected through diffraction gratings illuminating the detector arrays. By so doing the photon noise is reduced to that arising from the narrow band viewed by each detector, allowing the high sensitivity offered by state-of-the-art TES detectors to be fully utilized.  A generic grating module design is used for the four bands, with the infrared beam entering through an input slit and propagating via a collimator and a flat mirror to a high incidence grating. A mirror refocuses the dispersed signal onto the detector arrays. The practical size limit for the GM dictate the use of a high incidence grating, which allows for a more compact optics layout. As a result, the grating module is efficient for only a single polarization.

All elements in the grating modules are within a light-tight 1.8-K enclosure. Equally important, this enclosure  provides shielding for the  sensitive TES detectors and SQUID readouts against  electromagnetic interference (EMI). The Very Long Wavelength Grating module is shown in the right panel of Figure~\ref{Fig:SAFARIFPU} as an example of the GM design.

\begin{figure}
	\centering
	\includegraphics[width=\hsize]{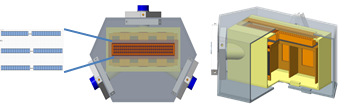}
	\caption{A generic 300mK SAFARI detector module. Left: schematic showing three lines of $1.5-F\lambda$ size detectors, separated by $4-F\lambda$. Centre: front view with horns in front of the TESs. The yellow box houses the 50-mK detector elements. The blue studs indicate the Kevlar suspension connection points, from 50 mK to 300 mK, to the 1.8-K structure. Right: cut-out view showing the cold readout AC biasing electronics, SQUIDs and LC filters.
	}
	\label{Fig:SAFARIFPA}
\end{figure}

SAFARI employs the latest generation of ultra-sensitive TES to detect the incoming photons \citep{2016SPIE.9914E..0BK,2016SPIE.9914E..08A,2016JLTP..184...52S,2012SPIE.8452E..0AG,2012SPIE.8452E..0GB}. TES bolometers fabricated at SRON \citep{2016JLTP..184...60R} have already demonstrated the required NEP \citep{2016SPIE.9914E..0BK}, as well as the required optical efficiency \citep{2016SPIE.9914E..08A}. To achieve their performance, the detectors, their readout and first stage amplifiers must be operated at 50-mK, which requires that both the sensors and the cold electronics be cooled and thermally isolated from the 1.8-K environment of the GM. This is achieved in two steps, with the TES and cold readout electronics in a 50-mK structure suspended, using Kevlar wires, from a 300-mK enclosure (the grey box in Figure~\ref{Fig:SAFARIFPA}). The 300-mK enclosure, providing additional radiation shielding, is again suspended using Kevlar from the 1.8-K Grating Module enclosure. The signal is coupled to the TES using $1.5F\lambda$ horns. There are three parallel linear TES arrays in the Focal Plane Assembly (FPA), thus constituting at each wavelength three separate spatial pixels, to provide background reference measurements for point source observations and redundancy against individual failures in TES sub-arrays.  The three spatial pixels also provide a limited imaging capability.

With some 3500 individual TES detectors  in  SAFARI system a detector readout multiplexing scheme is essential to limit the amount of wiring between the FPU and the warm electronics (Warm Front End Electronics, Detector Control Unit and Instrument Control Unit - WFEE, DCU, ICU) on the spacecraft SVM. A Frequency Domain Multiplexing (FDM) readout scheme will be employed in which each of the detectors in one multiplex channel is AC biased at a different frequency. The combined signals of the detectors in one channel are de-multiplexed and detected in the backend DCU. Multiplexing 160 detectors per channel, the designed flight configuration,  has already been demonstrated in a laboratory setup at SRON \citep{2016SPIE.9914E..1CH}.

To cool the TES detectors to their 50 mK operating temperature, a dedicated hybrid ADR/Helium sorption cooler is used. The cooler design builds on heritage from the Herschel and Planck missions. A full design has already been carried out for SAFARI, leading to the construction of an Engineering Model. Tests with the  Engineering Model show that the unit will be able to provide the required level of cooling with a $\sim$80\% duty cycle \citep{2014Cryo...64..213D}. The same cooler design will be used both for SAFARI and POL (see section~\ref{sec:POL}).

\subsubsection{The high resolution mode optics - the Martin-Puplett Interferometer}

\begin{figure}
	\centering
	\includegraphics[width=0.5\hsize]{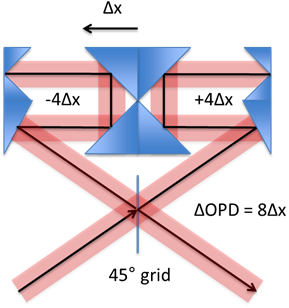}
	\caption{A Martin-Puplett interferometer: a linearly polarised input signal is divided over two arms of the interferometer using a 45$^{\rm o}$~ grid.  In both arms the beam goes via a flat mirror to a moving and a fixed rooftop and back, thus rotating the polarization three  times by 90$^{\rm o}$. This 270$^{\rm o}$~ rotated signal from the left arm is transmitted through the grid while from the right arm it is reflected, allowing the recombined beams to interfere. By moving the central rooftop mirrors over a distance $\Delta$x an optical path length difference of 8$\Delta$x is created between the two arms. The interference pattern, encoded in the polarization of the output signal, can be recorded by the grating module, due to its inherently linear polarization. }
	\label{Fig:SAFARIMPI}
\end{figure}

In the high-resolution SAFARI/HR mode the infrared beam is passes through a Martin-Puplett interferometer (Figure~\ref{Fig:SAFARIMPI},  see also \citet*{MartinPuplett}) which imposes a modulation on all wavelengths entering SAFARI. The resulting interference that occurs between the two beams of the interferometer is then distributed to the four GM (Figure~\ref{Fig:SAFARIFPU}) and post dispersed by the corresponding grating onto the detectors.

When the interferometer is scanned over its full optical displacement each of the detectors will measure a high resolution interferogram convolved with the grating response function for that particular detector. Upon Fourier transformation, an individual interferogram produces a small bandwidth, high resolution spectrum. By combining the spectra from individual detectors a full spectrum at high resolution is obtained. In the current design a mechanical displacement of about 3 cm is envisaged, leading, with a folding factor of 8, to a maximum optical displacement of 25 cm. A short section of the mechanism stroke must be devoted to a short double-sided optical path difference measurement to enable phase correction of the interferogram through accurate identification of the zero path difference position.  The available Optical Path Difference yields spectra with a resolving power ranging from $R\sim$1500 at 230 $\mu$m to as high as $R\sim$11000 at the shortest wavelength of 34 $\mu$m.

\subsubsection{Observing with SAFARI}

SAFARI will have a number of observing modes. Intrinsic to the design is instantaneous access to the full 34-230 $\mu$m wavelength domain. Thus in any mode, point source or mapping, low or high resolution,  a full spectrum will be obtained. The basic modes will be optimised for maximum efficiency for point source spectroscopy. In the $R\sim$300 grating mode point source spectra will be obtained using the BSM to chop between the source and a background off-source position, with the difference between the two giving a direct measure of the source flux. By chopping over a distance corresponding to the offset between the three pixel rows in the detector units there will always be one pixel on-source and two off-source, so that there will be  no time penalty for the background chopping.  In the SAFARI/HR mode chopping is not utilised because the scanning of the Martin-Puplett stage provides the modulation needed to correct for instrument drifts. Mapping modes are implemented using the BSM, providing a way to efficiently and flexibly cover small areas (<2') without requiring spacecraft repointing. For larger maps, on-the-fly  mapping modes will be implemented in which the spacecraft slowly `paints the sky' while the spectrometer continuously records data.

\subsection{The Mid-Infrared instrument SMI} 

The SMI mid-IR spectrometer/camera is designed to cover the wavelength range from 12 to 36 $\mu$m with both imaging and spectroscopic capabilities. The instrument employs four separate channels: the low-resolution spectroscopy function SMI/LR, the broad-band imaging function SMI/CAM, the mid-resolution spectroscopy function SMI/MR, and the high-resolution spectroscopy function SMI/HR. The prime science drivers for these functions are high-speed PAH spectral mapping of galaxies at $z > 0.5$ with SMI/LR, wide-area surveys of obscured AGNs and starburst galaxies at $z > 3-5$ with SMI/CAM, and velocity-resolved spectroscopy of gases in protoplanetary discs with SMI/HR. Complementary to these specific functions, SMI/MR provides more versatile spectroscopic functions, bridging the gap between JWST/MIRI \citep{2010AAS...21543904R} and SAFARI.

\begin{figure}
	\centering
	\includegraphics[width=\hsize]{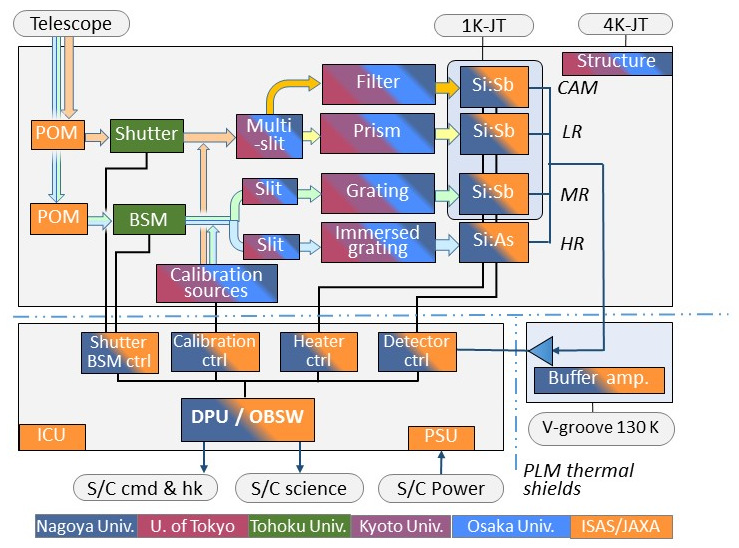}
	\caption{The SMI functional block diagram. The top half of the diagram represents the cold focal plane unit with its two sections. The top CAM/LR arm with the multislit/prism combination provides the combined fast wide field imaging and $R\sim$150 spectroscopy mode. The bottom arm with a beam steering mirror (BSM) forwarding the singanl to a slit/grating for the $R\sim$2000 MR mode, or to a slit/immersed grating for the $R\sim$28000 mode. The detector readout signals are sent through buffer amplifiers at the 130K level to the instrument Data Processing Unit (DPU) where the data are packaged for downlink. 
	}
	\label{Fig:SMIBlock}
\end{figure}

A functional block diagram for the SMI is shown in Figure~\ref{Fig:SMIBlock} with two main optics chains, one for SMI/LR-CAM and one for the SMI/MR-HR combination, each with their appropriate fore- and aft-optics. The only moving parts within SMI are the shutter in the SMI/LR-CAM chain and the beam steering mirror in the SMI/MR-HR chains. Table~\ref{tab:SMI} lists the SMI specifications.

\begin{table}
	\caption{SMI performance parameters.}
	\setlength{\tabcolsep}{2pt}
	\begin{tabular}{lccccl}
		\hline\hline
		Band 			
		& HR  			& MR  			& LR 			& CAM 
		\rule{0pt}{10pt} & \\ 
		\hline 
		$\lambda$ range 
		& 12-18   & 18-36  	& 17-36 	& 34  
		\rule{0pt}{10pt}  & $\mu$m \\
		$R$   				& 28000			& 2300-1300		& 50-120		&  5	&	\\ 
		\hline
		FoV
		& 4"$\times$1.7"&1'$\times$3.7"&10'$\times$3.7"& 10'$\times$12' 
		\rule{0pt}{10pt} & \\ 
		FWHM				& 2"			& 2.7"			& 2.7"			& 3.5"	&	\\ 
		Scale 		& 0.5			& 0.7			& 0.7			& 0.7	& "/pix.	\\
		\hline
		\multicolumn{5}{l}{Continuum sensitivity 5$\sigma$-1hr \rule{0pt}{10pt} }	& \\
		Point 		& 1500 			& 400  			& 50			& 13 	& $\mu$Jy	\\ 
		Diffuse  	& 	 			&   			& 0.05			& 0.05 	& MJy/sr \\ 
		\hline
		\multicolumn{5}{l}{Line sensitivity 5$\sigma$-1hr \rule{0pt}{10pt} } &	\\
		Point    	& 1.5			& 4 			&   8 & \multicolumn{2}{r}{ $10 ^{-20}$W/m$^2$\phantom{/sr} } \\ 
		Diffuse  	& 1.5			& 1 			&     & \multicolumn{2}{r}{$10^{-10}$W/m$^2$/sr} \\ 
		\hline
		Limit	& $\sim$20000	&$\sim$1000		& $\sim$20	& $\sim$1 & Jy \rule{0pt}{10pt}\\
		\hline\hline
	\end{tabular} 

	\label{tab:SMI}
\end{table}

SMI/LR is a multi-slit prism spectrometer with a wide field-of-view using four 10' long slits to execute low-resolution ($R = 50 - 120$) spectroscopic surveys with continuous coverage over the 17 - 36 $\mu$m wavelength domain. In  SMI/LR, a 10'$\times$12' slit viewer camera is implemented to accurately determine the positions of the slits on the sky for pointing reconstruction in creating spectral maps. This function provides an effective broad band imager centred at 34 $\mu$m - SMI/CAM. Both the spectrometer and the camera employ Si:Sb 1K$\times$1K detector arrays.  In the SMI/LR spectral mapping mode the multi-slit spectrometer and the camera are operated simultaneously, yielding both multi-object spectra from 17 to 36 $\mu$m as well as $R = 5$ deep images at 34 $\mu$m. The SMI/MR grating spectrometer covers the 18 -  36 $\mu$m wavelength range with a resolving power of $R = 1300-2300$.  The system employs a combination of an echelle grating and a cross-disperser. Like SMI/LR-CAM also this unit uses a 1K$\times$1K Si:Sb array for detection of the dispersed infrared beam. SMI/HR is a high resolution spectrometer covering the  12 - 18 $\mu$m wavelength range at  $R\sim$28000. This high resolution is achieved through the combination of an immersed grating and a cross-disperser using a ~4'' long slit. For this spectrometer a 1K$\times$1K Si:As array is used as detector.

\begin{figure*}[t]
	\begin{minipage}{0.65\hsize}
		\includegraphics[width=\hsize]{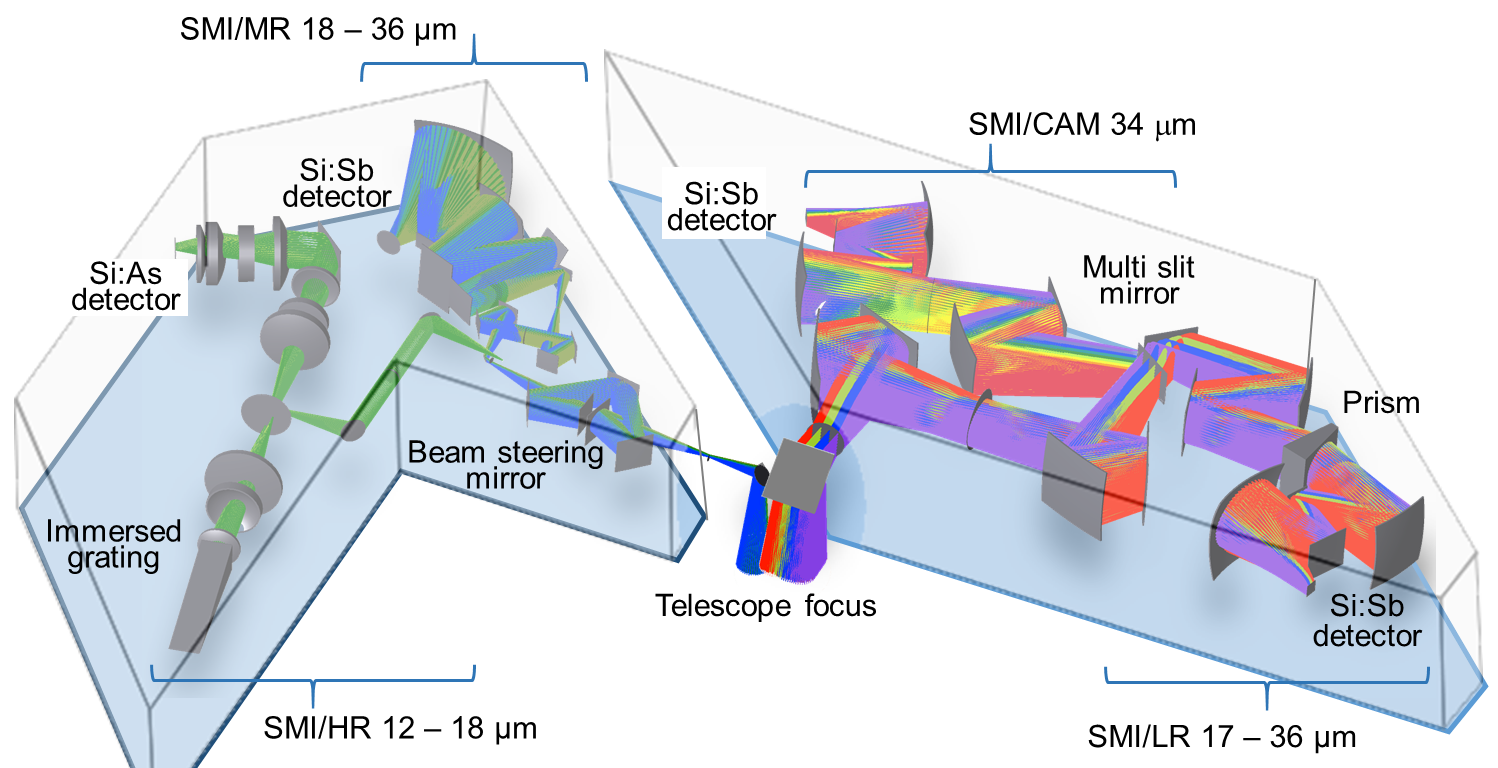}
	\end{minipage}
	\hfill
	\begin{minipage}{0.35\hsize}
		\centering
		\includegraphics[width=0.9\hsize]{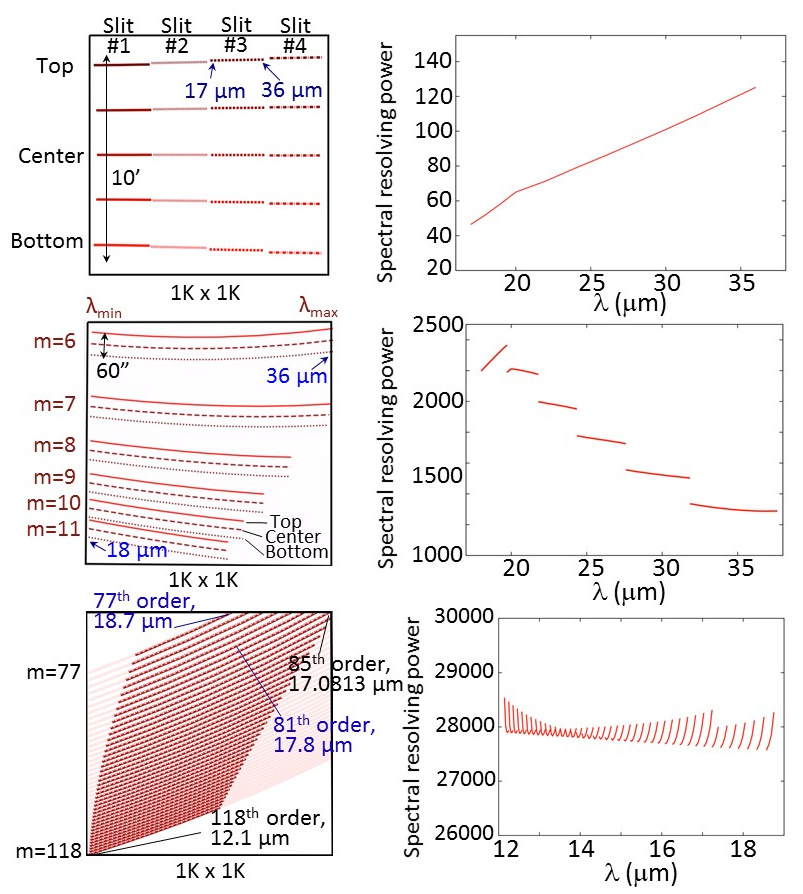}
	\end{minipage}
	\caption{Left: optical layout for SMI/MR-HR and SMI/LR with SMI/CAM. The color-coding of rays is based on angular positions in the fields-of-view. Right: spectral formats and spectral resolutions of SMI/LR, SMI/MR and SMI/HR (top to bottom).	}
	\label{Fig:SMIFPU}
\end{figure*}

\subsubsection{Optical design of SMI}

The optical layout of SMI is shown in Figure~\ref{Fig:SMIFPU}. The  design for SMI/LR-CAM and SMI/MR is based on reflective optics with aluminium free-form mirrors, while for the SMI/HR mainly refractive optics with lenses made of KRS-5, KBr, or CdTe are used. The fore-optics, optimised to remove the aberrations due to curvature and astigmatism in the incident beam, relay the telescope beam into the system. For SMI/LR-CAM (Figure~\ref{Fig:SMIFPU} right), a multi-slit plate with four slits, of 10' length and 3.7'' width, with a reflective surface is placed in the focal plane of the rear-optics entrance. The beam passing through the slits is directed to the spectrometer optics, while the beam reflected by the slit-plate is forwarded to the viewer channel optics. A KRS-5 prism in the pupil of the rear-optics of the spectrometer disperses the beam.  In front of the slit-viewer detector, a 34 $\mu$m  $R\sim$ 5 band-pass filter defines the bandpass for broad-band photometry. A wide field-of-view is realized with compact reflective optics using $6^{\rm th}$ order polynomial free-form mirrors. Diffraction-limited imaging performance is achieved (i.e., Strehl ratio > 0.8) for a 10'$\times$12' field-of-view at 17 $\mu$m for the spectrometer and at 30 $\mu$m for the viewer. The spectral resolution varies with wavelength from $R\sim$50 at 17 $\mu$m to $\sim$120 at 36 $\mu$m with slight (<10\%) differences between slits and positions within a slit (see Figure~\ref{Fig:SMIFPU} right, top panel).

As is the case for SMI/LR-CAM, the SMI/MR and SMI/HR share  fore-optics, including a beam-steering mirror. In combination with the telescope step-scan mode, for SMI/MR the beam steering mirror provides access to a $\sim$2'$\times$2' on-sky area. In the aft-optics, the beam is routed into either the SMI/MR or SMI/HR channels. SMI/MR has a long slit of 60" length and 3.7'' width. The beam passes through this slit and collimating optics, and is subsequently  dispersed by an Echelle grating combined with a cross-disperser. The resulting 18 -- 36 $\mu$m spectrum, spread over six different orders from m=6 to 11 (middle panel of Figure~\ref{Fig:SMIFPU} right panel).

SMI/HR has short, 4'' length, 1.7'' wide slit. It employs a CdZnTe immersed grating and a cross-disperser to disperse the signal from the slit, and collimating optics before the beam reaches the Si:As array. One part of the spectrum is obtained in 34 high orders (85th to 118th) covering 12.14 $\mu$m to 17.08 $\mu$m, and a second part in 8 lower orders (77th to 84th) partly covering the 17.08 $\mu$m to 18.75 $\mu$m range (Figure~\ref{Fig:SMIFPU}, right bottom panel).

\begin{figure}
	\centering
	\includegraphics[width=0.7\hsize]{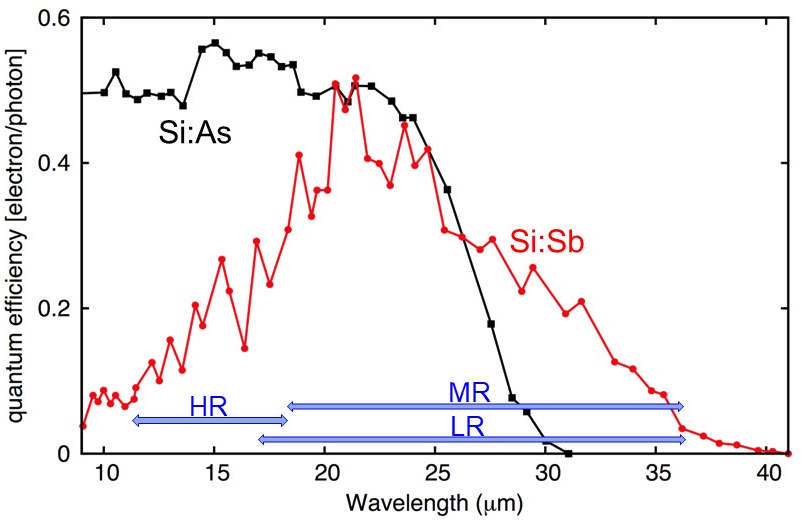}
	\caption{Quantum Efficiency for Si:As (JWST/MIRI; Ressler et al. 2008) and Si:Sb (test model for SMI; Khalap et al. 2012) arrays that are currently available. The SMI wavelength range is shown in blue.
	}
	\label{Fig:SMIDetectorPerformance}
\end{figure}

\subsubsection{The SMI detector arrays}

SMI employs two kinds of photoconductor arrays, Si:Sb 1K$\times$1K and Si:As 1K$\times$1K detectors.  Figure~\ref{Fig:SMIDetectorPerformance} shows the  quantum efficiencies that has been achieved by these types of detectors as a function of the wavelength. To achieve this performance the detectors must be operated at low temperatures: <2.0 K for Si:Sb, and <5.0 K for Si:As. Three Si:Sb arrays are used to cover wavelengths >17 $\mu$m (SMI/LR-CAM and SMI/MR) and one Si:As array for  wavelengths below 18 $\mu$m (SMI/HR). Si:As has heritage from previous space missions such as AKARI/IRC, Spitzer/IRAC and IRS, WISE and JWST/MIRI. To date Si:Sb has been used only for Spitzer/IRS with a 128$\times$128 array format.  Development is planned for the Si:Sb arrays towards 1K$\times$1K arrays with improved quantum efficiency at longer wavelengths.

\subsubsection{Observing with SMI}

SMI has four nominal  modes of observation: a staring mode, an SMI/LR mapping mode, an SMI/LR-CAM survey mode,  an SMI/MR mapping mode. The staring mode is used for targeted spectroscopy of point sources with SMI/LR-CAM, SMI/MR or SMI/HR. In this mode for SMI/MR and SMI/HR, dithering or mapping of a small area is possible using the beam-steering mirror. The SMI/LR mapping mode is used to perform slit scanning spectroscopy with SMI/LR to generate a full 10' x 12' spectral map with 90 scans each separated by a step of 2". The survey mode is used for wide-area surveys with SMI/LR-CAM in which the spacecraft is scanning with steps of $\sim$10' of the 10' x 12' field. The SMI/MR mapping mode is used for spectral mapping of extended sources by combining raster step scans and beam steering mirror movement. 

\subsection{A camera/polarimeter - POL}
\label{sec:POL}

The prime science driver for a far infrared polarimetric imaging function is the mapping of polarization in dust filaments in our Galaxy, requiring a high dynamic range both in spatial scales and flux density.  Efficient mapping requires an instantaneous field of view which is as large as possible and viewed simultaneously in different wavelength bands. Thus detectors are required that offer good sensitivity at faint flux levels, but are not affected by high flux levels.  The wavelength bands for POL are defined by the need to observe filaments on both sides of their peak emission, and centred around 100$\mu$m, 200$\mu$m, and 350$\mu$m.  For efficient polarimetry, polarizing detectors are used.  The specifications of the instrument are summarized in Table~\ref{tab:POL}. The adopted detector sensitivity of $3\times 10^{-18}$ W/$\surd$Hz can be achieved with detectors which were originally developed for an earlier incarnation of the SAFARI spectrometer. 

\begin{figure}
	\centering
	\includegraphics[width=\hsize]{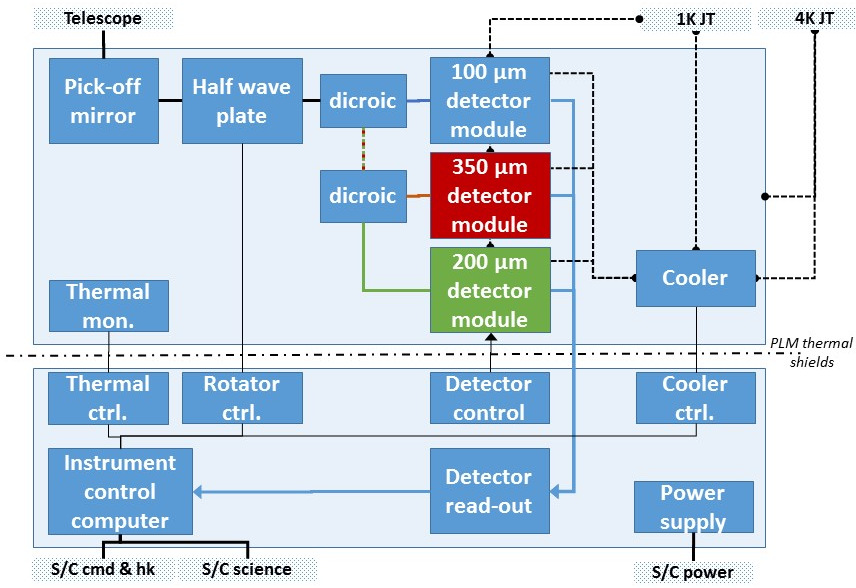}
	\caption{POL functional block diagram, implementing individual direction of three  wavelength bands onto a common focal plane assembly.  An achromatic half-wave plate in the common part of the optics train is used by all bands.}
	\label{Fig:POLBlock}
\end{figure}

\subsubsection{The POL optics}

\begin{figure*}
	\centering
	\begin{minipage}{0.25\hsize}
		\includegraphics[width=\linewidth]{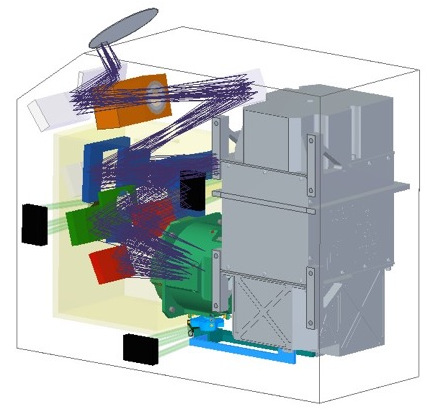}
	\end{minipage}
	\begin{minipage}{0.25\hsize}
	\centering
	\includegraphics[width=0.8\linewidth]{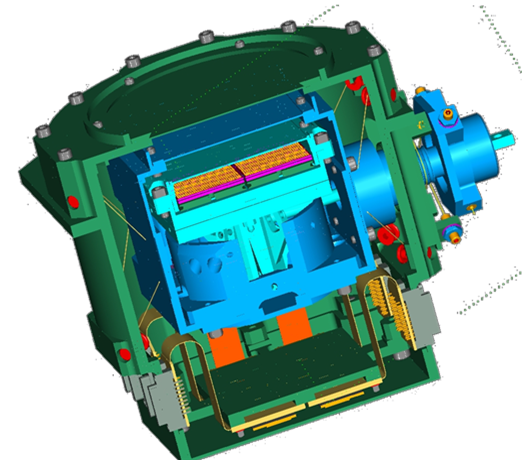}
	\end{minipage}
	\begin{minipage}{0.25\hsize}
		\centering
		\includegraphics[width=0.7\linewidth]{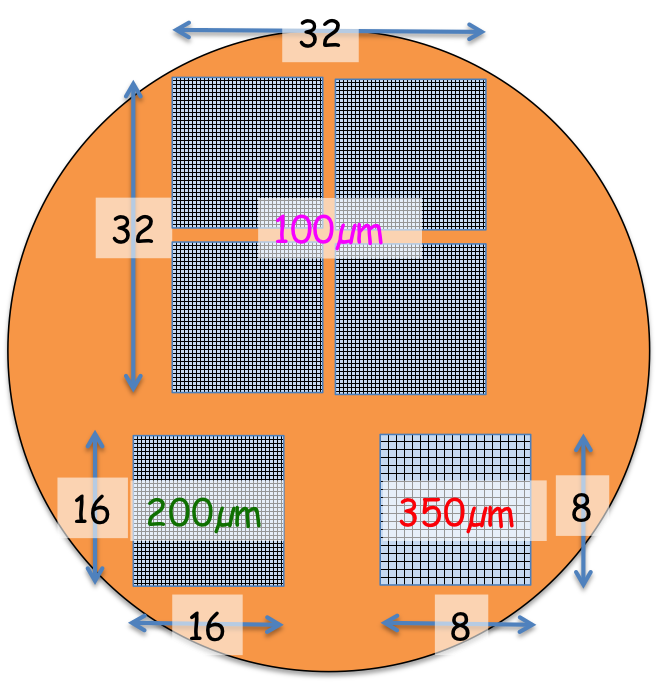}
	\end{minipage}
	\begin{minipage}{0.14\hsize}
	\centering
	\includegraphics[width=\linewidth]{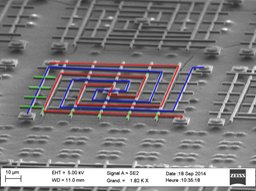}
	\includegraphics[width=\linewidth]{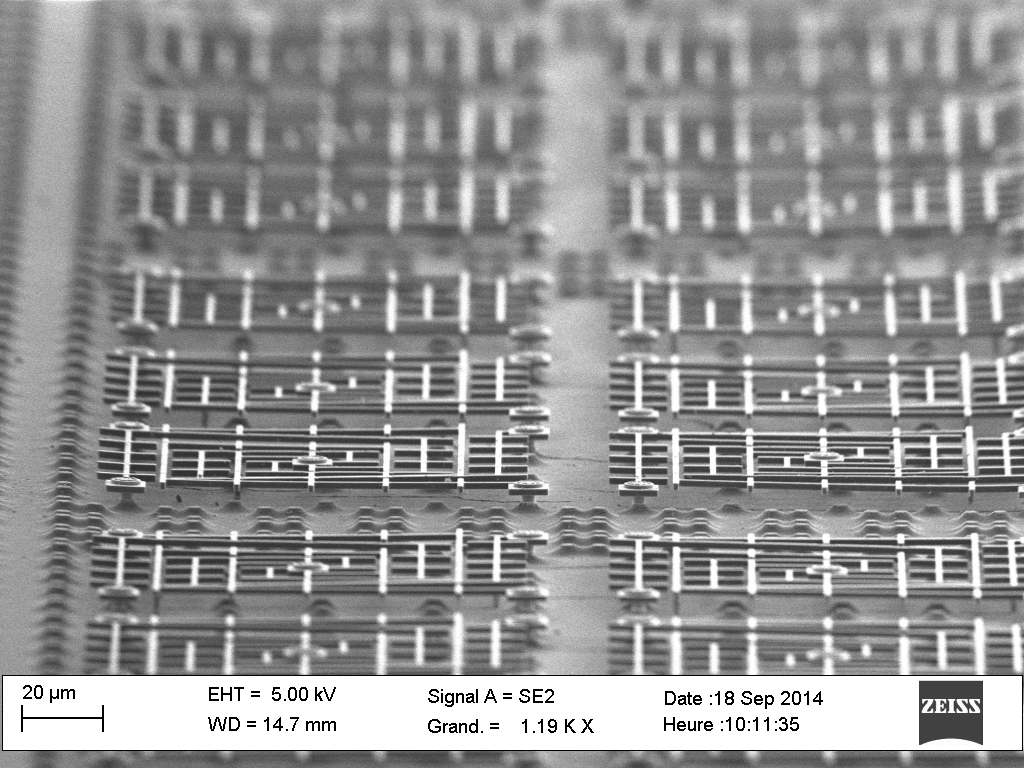}
	\end{minipage}
    \caption{POL Focal Plane Unit, detector assembly and detectors.   Left: the three band layout of POL. Right: the FPA structure - light blue indicates the 50-mK stage, dark blue the 300-mK stage and green the 1.8-K stage. The two inner stages are suspended by Kevlar wires.  \changed{ Right top: a single POL pixel showing an interlaced grid of small metallic vertical and horizontal polarization absorbers on a Si substrate. These 'antennae' absorb one polarization of the  incoming radation and the power dissipation results in temperature changes in the support structure. These changes are is sensed using the four spiral arm thermistor leads - two of the leads couple to the structure supporting the horizontal polarization antennae, and two to the vertical polarization. Right bottom: an array of the POL pixels.} }
	\label{Fig:POLFPU}
\end{figure*}

The instrument optical layout is shown in Figure~\ref{Fig:POLFPU}, it employs common entrance optics for the three bands, providing field- and aperture stops for stray light suppression as well as a pupil image for the half-wave plate, followed by individual branches for the three wavelength bands.  The half-wave plate is an achromatic design, providing nearly constant phase shift across the entire wavelength range.  The plate is operated at 4.8 K and it is immediately followed by a 1.8 K Lyot stop.  All further elements are inside a 1.8 K enclosure.  After the intermediate focus/field stop, light is re-collimated and split into the three wavelength bands by means of two dichroic filters. The collimated beams in the three separate bands are re-focused onto the three detector arrays with different magnification, to match physical pixel sizes and optical point spread functions at the band centre wavelengths.  A boundary condition for the optics design has been to align the three detector focal planes close enough in position and orientation to allow their integration in one common focal plane assembly (FPA) at 50 mK.  Band-defining filters are mounted on the 300-mK and 50-mK levels of the focal plane assembly.  To optimise between sensitivity, mapping speed, and resolution, a 1.22/2$F\lambda$ pixel size is used.

\begin{table}
	\caption{POL performance parameters}
	\setlength{\tabcolsep}{3pt}
	\centering
	\begin{tabular}{lccc}
		\hline\hline
		Band 			& 100 $\mu$m  	& 200 $\mu$m  		& 350 $\mu$m \rule{0pt}{10pt} \\ 
		\hline 
		$\lambda$ range & 75-125 $\mu$m	& 150-250 $\mu$m	& 280-420 $\mu$m \rule{0pt}{10pt} \\ 
		Array size 		& 32$\times$32  & 16$\times$16 		& 8$\times$8 		\\ 
		Pixel size 		& 5"$\times$5" 	& 10"$\times$10" 	& 20"$\times$20" 	\\ 
		FWHM			& 9"			& 18"				& 32"				\\
		\hline 
		\multicolumn{4}{l}{Point source sensitivity 2.5'$\times$2.5' 5$\sigma$-1hr \rule{0pt}{10pt} } \\
		Unpol.		& 21 $\mu$Jy	& 42 $\mu$Jy		& 85 $\mu$Jy		\\
		 Q, U 	& 30 $\mu$Jy	& 60 $\mu$Jy 		& 120 $\mu$Jy		\\
		\hline 
		\multicolumn{4}{l}{Point source sensitivity  1 deg$^2$ 5$\sigma$-10hr \rule{0pt}{10pt} } \\
		Unpol.			& 160 $\mu$Jy	& 320 $\mu$Jy		& 650 $\mu$Jy		\\
		 Q, U 	& 230 $\mu$Jy	& 460 $\mu$Jy		& 920 $\mu$Jy		\\
		\hline
		\multicolumn{4}{l}{Surface brightness sensitivity  1 deg$^2$ 5$\sigma$-10hr \rule{0pt}{10pt} } \\
		Unpol.			& 0.09 MJy/sr	& 0.045 MJy/sr		& 0.025 MJy/sr		\\
		Q, U 	& 2.5 MJy/sr	& 1.25 MJy/sr		& 0.7 MJy/sr		\\
		\hline\hline
	\end{tabular} 

	\label{tab:POL}
\end{table}

The instrument requires an actuator for the half-wave plate at the 4-K temperature level.  An electromagnetic motor design based on the Herschel/PACS \citep{2010A&A...518L...2P} filter wheels or the more compact motor used in FIFI-LS \citep{2014SPIE.9147E..2XK} on SOFIA \citep{2016SPIE.9973E..0IB} can be used.  To minimize dissipation during movements, an integrated wheel/motor design is foreseen, as the main dissipation occurs not in the motor coils but by mechanical friction in the bearings of the mechanism. The half-wave plate is rotated only between scans, and its operation will contribute marginally to the thermal load on the 4-K-JT stage.

\subsubsection{The POL detector assembly}

The detector assembly (see Figure~\ref{Fig:POLFPU}) holds three detector ensembles, one for each spectral band. The 100-$\mu$m band has four 16$\times$16 pixel modules butted in a 2$\times$2 configuration, the 200-$\mu$m band has one 16$\times$16 pixel module, and the 350-$\mu$m band one 8$\times$8 module. All modules have the same physical size, $\sim$20 mm, and the same field of view.

The detector assembly is a `Russian dol'' structure at 1.8 K with two suspended stages. The innermost level at 50 mK houses the six detector modules and is thermally linked to the 50-mK cryocooler tip by a light tight coaxial pipe. The 50-mK level is enclosed in a 300-mK box linked to the cryocooler by the same cooling pipe. POL will use the same cooler design as SAFARI. To prevent out-of-band stray light, the beam passes through filters at the entrance to each enclosure. 

To achieve sufficient dynamic range semiconductor bolometers are used, with heritage from the Herschel PACS bolometer arrays, redesigned to support polarization measurement (Figure~\ref{Fig:POLFPU}), and cooled to 50 mK to achieve the required sensitivity. These resistive sensors do not show any saturation with absorbed power, but they do suffer from a non-linear response. This non-linearity can be taken in account by applying a proper calibration over a wide range in incident power.

The bolometer detector consists of two suspended interlaced spirals, each sensitive to a single polarization component. The metallic absorbers (here dipole antennae matched to vacuum impedance) form a resonant cavity with a reflector on the readout circuit surface. The cavity is partially filled with a dielectric (SiO) to tailor the detector bandwidth. The frontend readout circuit also operates as a buffer stage output for the time domain multiplexing function, both operating at 50~mK. The multiplexing leads to a large reduction in the number of connections to the coldest stage, minimising the thermal load. The large difference in dynamic range between total power and polarization is addressed by the use of a Wheatstone bridge configuration. The three Stokes parameters ($I, Q$ and $U$) can in principle be retrieved simultaneously in the PSF by a `polka dot' configuration with every other detector rotated by $45\deg$. Nevertheless, a half-wave plate located at the instrument entrance is necessary to disentangle scene polarization from instrumental self-polarization.

\subsubsection{Observing with POL}

The foreseen operating mode of the detectors will produce a combined, total power and difference signal for two orthogonal polarizations.  For simple imaging, any mapping scheme can be used.  For polarimetry, the default observing mode will be scan maps, along two approximately orthogonal scan directions.  The scan map will be repeated with one or more different orientations of the half-wave plate.  The wave plate serves two purposes: it provides access to the $\pi /4$ and $3 \pi /4$ orientations and allows the removal of polarization effects introduced by the instrument by swapping the two orthogonal polarizations of the detector. To establish reliable end-to-end characterization of the polarization properties of the complete system (telescope plus instrument), calibration observations will be repeated at regular intervals during the mission.

\begin{figure}
	\centering
	\includegraphics[width=\hsize]{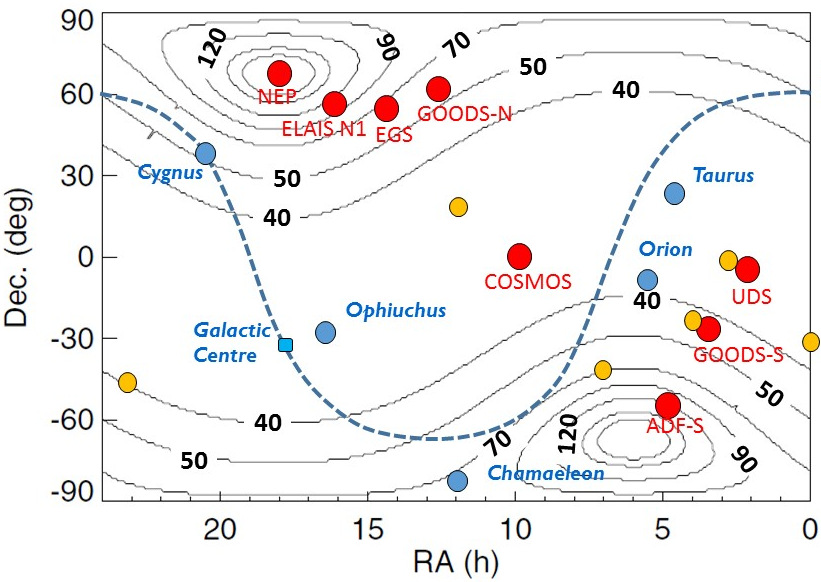}
	\caption{Sky visibility contours, in unit of days per year. Circles identify popular extragalactic survey fields (red), HST Frontier Fields (yellow), and Galactic star forming regions (blue).
	}
	\label{Fig:SPICASkyVisibility}
\end{figure}

\section{SPICA operations}  

\subsection{Launch and operations}

SPICA is envisaged to be launched into space on JAXA's next-generation H3 launch vehicle from the JAXA Tanegashima Space Centre. The satellite will be directed to a halo orbit around the Sun-Earth Libration point 2 (S-E L2) which provides a stable thermal environment. The orbit will give access to a 360$^{\rm o}$~ annulus with a width of about 16$^{\rm o}$~ on the sky, providing full sky access over a six-month period  (see Figure~\ref{Fig:SPICASkyVisibility}). Most the cosmological fields of interest have good visibility (over 40 days per year), deep observations of fields like COSMOS and UDS likely will require re-visits over multiple years. For galactic sources the visibility varies, unfortunately with  somewhat poorer visibilities for some of the prototypical sites of galactic star formation like Taurus and Ophiuchus.  The spacecraft will be operated in a 24-hour cycle autonomous operation. In a daily contact period, a new 24-hour schedule will be uploaded while instrument and spacecraft data stored in mass memory will be downloaded.

\begin{figure}[b]
	\centering
	\includegraphics[width=\hsize]{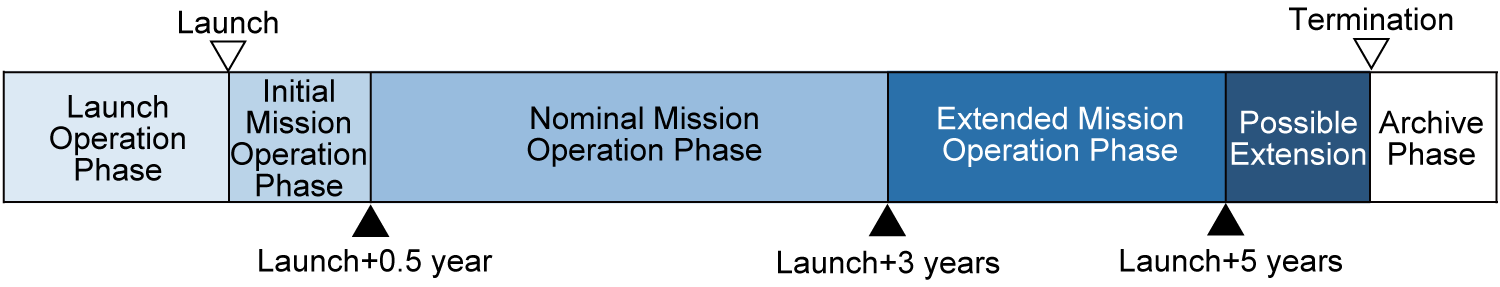}
	\caption{Operational phases of the SPICA mission.
	}
	\label{Fig:SPICAMissionPhases}
\end{figure}

The spacecraft will be launched at ambient temperature, and the payload will be cooled to the operating temperature in early mission operations. The lifetime required for the mission will be three years, with a goal of five years. Given the fact that mission lifetime does not depend on a limited cryogen supply, a further extension is quite conceivable; the cryogen-free design would allow extension of the mission lifetime beyond the nominal duration and is ultimately limited only by propellant and potential on-board hardware degradation. The different operational phases of the mission are indicated in Figure~\ref{Fig:SPICAMissionPhases}.

\subsection{Ground segment}

\begin{figure}[b]
	\centering
	\includegraphics[width=\hsize]{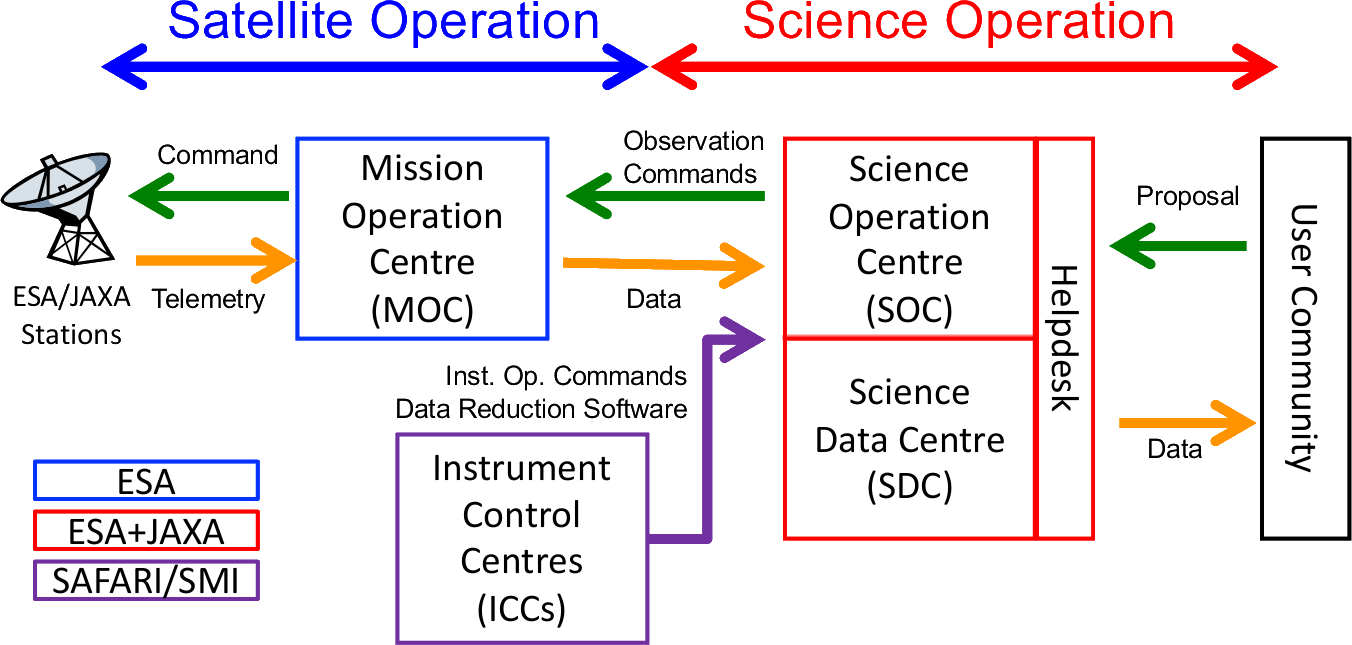}
	\caption{SPICA operational centres and information flow.
	}
	\label{Fig:SPICAGroundSegment}
\end{figure}

The ground segment will be designed to run the mission such that its scientific harvest is maximised. It consists of the Mission Operation Centre (MOC), the Science Operation Centre (SOC), the Science Data Centre (SDC) and the Instrument Control Centres (ICCs).  Figure~\ref{Fig:SPICAGroundSegment} shows these centres together with the main flows of information.

The MOC is responsible for all spacecraft operations including the execution of routine observations and contingency plans. It will monitor the health and safety of the spacecraft and instruments, and when needed to take corrective actions, and will generate and upload commands based on the observation plan input from the SOC and receive telemetry data. The MOC is expected to be operated in Europe (by ESA). The SOC will be in charge of the science operation of SPICA, i.e. the handling of observing proposals, generating schedules of approved science and calibration observations for the MOC to generate and upload daily commands, and handling the downlinked science data. The SDC will develop and maintain the data reduction toolkit/pipeline software together with the ICCs. The data will be processed and archived by the SDC, and made available to the ICCs and science users. The SOC and SDC will jointly operate a help-desk as a unique contact point for science users. The SOC and SDC will be established in the framework of collaboration between ESA and JAXA.

The ICCs, under the responsibility of the instrument teams, will work together with the SOC and the SDC to ensure that the focal-plane instruments are well calibrated and optimally operated. They will monitor instrument health, define and analyse calibration observations, and will also be responsible for developing and maintaining the scientific data reduction software.

\subsubsection{Operation scheme}

As soon as the fundamental spacecraft functions have been successfully verified, functional checkout of the focal plane instruments will start followed by the scientific performance verification (PV) phase. When warranted by the PV results, observing programmes may need to be adjusted to account for the established in-flight performance. The spacecraft will carry out observations autonomously, according to a timeline of commands uploaded from the ground. The baseline operational mode foresees no parallel mode operations - at any one time only a single instrument will be active.	

Data taken by the instruments will be made available to the data owners for scientific analysis. In the routine phase, observation data will be delivered to the users within one month after successful execution of the observation. Calibration and data reduction software will be updated regularly during the operation. Major updates will be made available every 0.5-1.0 yr. All data will be re-processed by the SDC using the new pipeline.

\subsection{Science programmes}

SPICA is to be operated as an international observatory to accomplish the mission science goals as well as to execute science programmes proposed by the wider community. Two categories of observing programmes are foreseen; Key Programmes (KP: significant, consistent and systematic programmes to carry out the mission's prime science goals), and General Programmes (GP, all other observing programmes). Observing time will be divided into Guaranteed time (GT: reserved for instrument consortia and other groups involved in developing the mission),  Open Time (OT: open to the world wide community) and Director's Discretionary Time (DDT). For obvious reasons, a major part of the GT (> 70\%) shall be defined as KP.

Any programme proposed by consortium members (GT) or coming from the world-wide community (OT) could be a Key Programme or a General Programme, depending on its goals and/or nature. A Time Allocation Committee (TAC) will review all observing time proposals and propose prioritized time allocations to the Scientific Advisory Board (SAB) on the basis of scientific merit. The SAB will subsequently ratify the observation programme with the priorities proposed by the TAC. It will also be the role of the SAB to define what should be the proportion of KP in the OT.

\begin{figure}
	\centering
	\includegraphics[width=0.6\hsize]{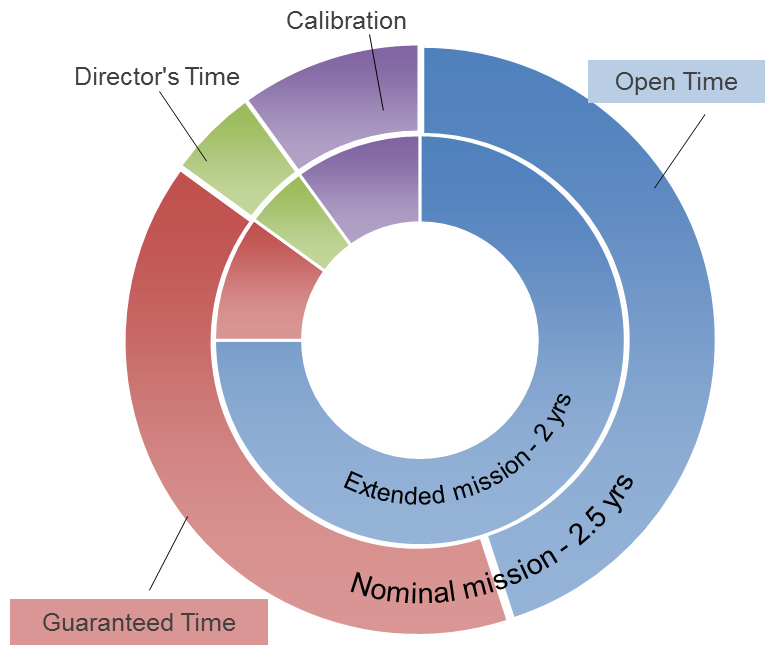}
	\caption{Division of  observing time over different programme categories.
	}
	\label{Fig:SPICAObservingTime}
\end{figure}

In the nominal 3-year mission the first six months will be used for cooling the telescope, and verification and validation of the observatory and its instruments. OT proposals will be assigned the largest fraction of the remaining 2.5 years, while a more modest part (<40\%) will be dedicated to GT and a small (<5\%) part will be reserved as DDT. In a possible mission extension, the proportion of GT will be small (< 10\%), the DDT will remain the same (< 5\%) and most of the time will be OT open to the world-wide community. Figure~\ref{Fig:SPICAObservingTime} illustrates this concept for the time division over programme categories.

The SPICA project will be responsible for Level-2 processing of the data, i.e., correction for instrument-specific features and calibration to make the data ready for scientific analysis. All science data will be archived, and made publicly available after a proprietary period of one year.

\section{Conclusions}

The joint European-Japanese infrared space observatory SPICA will provide a major step in mid- and far- infrared astronomical research capabilities after the Herschel mission.  To minimise the background noise level SPICA will employ a 2.5 meter telescope cooled to below 8 K. As a result the detectors are no longer affected by the thermal radiation coming from the mirror itself, allowing the ultrasensitive SPICA instruments to detect infrared sources over  two orders of magnitude weaker than would have been possible with previous infrared space observatories. 

The instrument complement foreseen for SPICA will provide extremely sensitive specroscopic capabilities in the 12 to 230 $\mu$m domain with various modes in resolving power - ranging from extremely sensitive  $R=300$ spectroscopy instantaneously covering the full 34 to 230 $\mu$m band to a high resolution $R\sim 28000$ mode between 12 and 17 $\mu$m.  In addition large field of view sensitve imaging photometry at 34$\mu$m and imaging polarimetry at 100, 200 and 350 $\mu$m  is provided. The observatory cryogenic system is based on a combination of passive cooling and mechanical cryocoolers, making the operational lifetime independent of liquid cryogens. 

With SPICA Astronomers will e.g.  be able to study the process of galaxy evolution over cosmic time in much more detail and out to much larger redshift than was possible before. Furthermore the observatory will allow detailed investigatiuons into the process of planet formation, as well as the study of the role of the galactic magnetic field in starformation in dust filaments.

\begin{acknowledgements}
	
	This paper is dedicated to the memory of Bruce Swinyard and Roel Gathier. Bruce initiated the SPICA project in Europe, but sadly died on 22 May 2015 at the age of 52. He was ISO-LWS calibration scientist, Herschel-SPIRE instrument scientist, first European PI of SPICA and first design lead of SAFARI. Roel was managing director of SRON  until early 2016 when he died after a short sickbed on 14 March 2016. Roel also was head of the Dutch delegation in the Science Programme Committee (SPC) and later SPC chairman. He was crucial in giving and generating support for the SPICA mission in the Netherlands, but also throughout Europe, Japan and north America.

\changed{
The SAFARI project in the Netherlands is financially supported through NWO grant for Large Scale Scientific Infrastructure nr 184.032.209 and NWO PIPP grant NWOPI 11004. The TES detector development work for SAFARI has received support from the European Space Agency (ESA) TRP program (contract number 22359/09/NL/CP).

FN and JTR acknowledge financial support through Spanish grant ESP2015-65597-C4-1-R (MINECO/FEDER).
}

\end{acknowledgements}

\begin{appendix}

\section*{Affiliations}

         \affil{$^1$SRON Netherlands Institute for Space Research, Postbus 800, 9700 AV, Groningen, The Netherlands\label{inst:SRON} }
         \affil{$^2$Kapteyn Astronomical Institute, University of Groningen, Postbus 800, 9700 AV, Groningen, The Netherlands \label{inst:Kapteyn} }
         \affil{$^3$Department of Earth and Space Science, Graduate School of Science,  Osaka University, 1-1 Machikaneyama, Toyonaka, Osaka 560-0043, Japan \label{inst:Osaka} }
         \affil{$^4$Institute of Space Astronautical Science, Japan Aerospace Exploration Agency, 3-1-1 Yoshinodai, Chuo-ku, Sagamihara, Kanagawa 252-5210, Japan \label{inst:ISAS} }
         \affil{$^5$Spitzer Science Center, California Institute of Technology, MC 314-6, Pasadena, CA 91125, USA \label{inst:SSC} }
         \affil{$^6$Department of Astronomy, University of Geneva, ch. d'Ecogia 16, CH-1290 Versoix, Switzerland \label{inst:ISDC} }
         \affil{$^7$JPL, USA \label{inst:JPL} }
         \affil{$^8$Department of Astronomy, California Institute of Technology, 1200 E. California Blvd., Pasadena, CA 91125, USA \label{inst:Caltech} }
         \affil{$^9$Department of Earth Science and Astronomy, Graduate School of Arts and Sciences, University of Tokyo, 3-8-1 Komaba, Meguro-ku, Tokyo 153-8902, Japan \label{inst:Tokyo-DESA} }
         \affil{$^{10}$IRAP, Universit\'e de Toulouse, CNRS, CNES, UPS, (Toulouse), France \label{inst:IRAP} }
         \affil{$^{11}$Istituto di Astrofisica e Planetologia Spaziali, INAF-IAPS, Via Fosso del Cavaliere 100, I-00133 Roma, Italy \label{inst:INAF} }
         \affil{$^{12}$Cardiff school of Physics and Astronomy, The Parade, Cardiff, CF24 3AA, UK \label{inst:Cardiff} }
         \affil{$^{13}$Graduate School of Science, Nagoya University, Furo-cho, Chikusa-ku, Nagoya 464-8602, Japan \label{inst:Nagoya} }
         \affil{$^{14}$Academia Sinica, Institute of Astronomy and Astrophysics, PO Box 23-141, Taipei 10617, China, Taipei \label{inst:ASIAA} }
         \affil{$^{15}$University of Vienna, Department of Astrophysics, Türkenschanzstrasse 17, 1180, Wien, Austria \label{inst:Vienna} }
         \affil{$^{16}$Institute of Astronomy,  University of Tokyo, 2-21-1 Osawa, Mitaka, Tokyo 181-0015, Japan \label{inst:Tokyo-IofA} }
         \affil{$^{17}$Max Planck Institute for Astronomy Königstuhl 17, Heidelberg D-69117, Germany \label{inst:MPIA} }
         \affil{$^{18}$Stockholm Observatory, Stockholm University, AlbaNova University Center, 106 91, Stockholm, Sweden \label{inst:Stockholm} }
         \affil{$^{19}$Laboratoire AIM, CEA/IRFU/Service d'Astrophysique, Universit\'e Paris Diderot, Bat. 709, F-91191 Gif-sur-Yvette, France \label{inst:CEA-AIM} }
         \affil{$^{20}$Centro de Astrobiologia (CSIC-INTA), Ctra de Torrejón a Ajalvir, km 4, 28850, Torrejon de Ardoz (Madrid), Spain \label{inst:CAB} }
         \affil{$^{21}$Institute for Space Imaging Science, Department of Physics and Astronomy, University of Lethbridge, Lethbridge, Alberta, T1K 3M4, Canada \label{inst:Lethbridge} }
         \affil{$^{22}$Department of Astronomy, Graduate School of Science, University of Tokyo, 7-3-1 Hongo, Bunkyo-ku, Tokyo 113-0033, Japan \label{inst:Tokyo-DofA} }
         \affil{$^{23}$Max-Planck-Institut f\"{u}r extraterrestrische Physik, Postfach 1312, D-85741, Garching, Germany. \label{inst:Poglitch} }
         \affil{$^{24}$CEA Saclay, Service d'Astrophysique, L'Orme des Merisiers Bat 709, BP 2, 91191 Gif-sur-Yvette Cedex, France \label{inst:CEASaclay} }
         \affil{$^{25}$Instituut voor Sterrenkunde, KU Leuven, Celestijnenlaan 200 D, 3001, Leuven, Belgium \label{inst:KUL} }
         \changed{
         \affil{$^{26}$INTA Instituto Nacional de T\'ecnica Aeroespacial, Ctra de Torrej\'on a Ajalvir, km 4, 28850, Torrejon de Ardoz (Madrid), Spain \label{inst:INTA} }
         \affil{$^{27}$Cavendish Laboratory, JJ Thomson Avenue, Cambridge CB3 0HE UK  \label{inst:Cavendish} }
         \affil{$^{28}$Univ. Grenoble Alpes, CEA, INAC-SBT, 38000 Grenoble, France  \label{inst:CEA-INAC} }
         \affil{$^{29}$Research and Development Directorate, Japan Aerospace Exploration Agency,  2-1-1, Sengen, Tsukuba, Ibaraki 305-8505, Japan  \label{inst:JAXA-RDD} }
                       }

\end{appendix}

\bibliographystyle{pasa-mnras} 
\bibliography{SPICAReferences}

\begin{thebibliography}{}
\makeatletter
\relax
\def\mn@urlcharsother{\let\do\@makeother \do\$\do\&\do\#\do\^\do\_\do\%\do\~}
\definecolor{darkblue}{rgb}{0,0,0.597656}
\def\mndoi{\begingroup\mn@urlcharsother \@ifnextchar [ {\mndoi@} {\mndoi@[]}}
\def\mndoi@[#1]#2{\def\@tempa{#1}\ifx\@tempa\@empty \href
  {http://dx.doi.org/#2} {\textcolor{darkblue}{doi:#2}}\else \href
  {http://dx.doi.org/#2} {\textcolor{darkblue}{#1}}\fi \endgroup}
\def\mn@eprint#1#2{\mn@eprint@#1:#2::\@nil}
\def\mn@eprint@arXiv#1{\href {http://arxiv.org/abs/#1} {{\tt arXiv:#1}}}
\def\mn@eprint@dblp#1{\href {http://dblp.uni-trier.de/rec/bibtex/#1.xml}
  {dblp:#1}}
\def\mn@eprint@#1:#2:#3:#4\@nil{\def\@tempa {#1}\def\@tempb {#2}\def\@tempc
  {#3}\ifx \@tempc \@empty \let \@tempc \@tempb \let \@tempb \@tempa \fi \ifx
  \@tempb \@empty \def\@tempb {arXiv}\fi \@ifundefined
  {mn@eprint@\@tempb}{\@tempb:\@tempc}{\expandafter \expandafter \csname
  mn@eprint@\@tempb\endcsname \expandafter{\@tempc}}}

\bibitem[\protect\citeauthoryear{{Andr\'{e}}}{{Andr\'{e}}}{2017}]{WP8-Andre}
{Andr\'{e}} P.,  2017, \pasa, in prep

\bibitem[\protect\citeauthoryear{{Audley}, {de Lange}, {Gao}, {Khosropanah},
  {Hijmering}  \& {Ridder}}{{Audley} et~al.}{2016}]{2016SPIE.9914E..08A}
{Audley} M.~D.,  {de Lange} G.,  {Gao} J.-R.,  {Khosropanah} P.,  {Hijmering}
  R.,   {Ridder} M.~L.,  2016, in Millimeter, Submillimeter, and Far-Infrared
  Detectors and Instrumentation for Astronomy VIII. p. 991408 (\mn@eprint
  {arXiv} {1608.06843}), \mndoi{10.1117/12.2231088}

\bibitem[\protect\citeauthoryear{{Becklin}, {Young}  \& {Savage}}{{Becklin}
  et~al.}{2016}]{2016SPIE.9973E..0IB}
{Becklin} E.~E.,  {Young} E.~T.,   {Savage} M.~L.,  2016, in Infrared Remote
  Sensing and Instrumentation XXIV. p. 99730I, \mndoi{10.1117/12.2238788}

\bibitem[\protect\citeauthoryear{{Beyer} et~al.,}{{Beyer}
  et~al.}{2012}]{2012SPIE.8452E..0GB}
{Beyer} A.~D.,  et~al., 2012, in Millimeter, Submillimeter, and Far-Infrared
  Detectors and Instrumentation for Astronomy VI. p. 84520G,
  \mndoi{10.1117/12.926326}

\bibitem[\protect\citeauthoryear{{Cohen}}{{Cohen}}{2016}]{2016APS..APR.E7002C}
{Cohen} J.,  2016, in APS April Meeting Abstracts.

\bibitem[\protect\citeauthoryear{{Dole} et~al.,}{{Dole}
  et~al.}{2006}]{2006A&A...451..417D}
{Dole} H.,  et~al., 2006, \mndoi [\aap] {10.1051/0004-6361:20054446}, \href
  {http://adsabs.harvard.edu/abs/2006A%26A...451..417D} {451, 417}

\bibitem[\protect\citeauthoryear{{Duband}, {Duval}  \& {Luchier}}{{Duband}
  et~al.}{2014}]{2014Cryo...64..213D}
{Duband} L.,  {Duval} J.~M.,   {Luchier} N.,  2014, \mndoi [Cryogenics]
  {10.1016/j.cryogenics.2014.02.008}, \href
  {http://adsabs.harvard.edu/abs/2014Cryo...64..213D} {64, 213}

\bibitem[\protect\citeauthoryear{{Egami} et~al.,}{{Egami}
  et~al.}{2017}]{WP6-Egami}
{Egami} E.,  et~al., 2017, in prep.

\bibitem[\protect\citeauthoryear{{Fern\'andez-Ontiveros}
  et~al.,}{{Fern\'andez-Ontiveros} et~al.}{2017}]{WP2-Fernandez-Onteveros}
{Fern\'andez-Ontiveros} J.,  et~al., 2017, \pasa, this issue

\bibitem[\protect\citeauthoryear{{Gaia Collaboration} et~al.,}{{Gaia
  Collaboration} et~al.}{2016}]{2016A&A...595A...1G}
{Gaia Collaboration} et~al., 2016, \mndoi [\aap] {10.1051/0004-6361/201629272},
  \href {http://adsabs.harvard.edu/abs/2016A%26A...595A...1G} {595, A1}

\bibitem[\protect\citeauthoryear{{Gardner} et~al.,}{{Gardner}
  et~al.}{2006}]{2006SSRv..123..485G}
{Gardner} J.~P.,  et~al., 2006, \mndoi [\ssr] {10.1007/s11214-006-8315-7},
  \href {http://adsabs.harvard.edu/abs/2006SSRv..123..485G} {123, 485}

\bibitem[\protect\citeauthoryear{{Goldie} et~al.,}{{Goldie}
  et~al.}{2012}]{2012SPIE.8452E..0AG}
{Goldie} D.~J.,  et~al., 2012, in Millimeter, Submillimeter, and Far-Infrared
  Detectors and Instrumentation for Astronomy VI. p. 84520A,
  \mndoi{10.1117/12.925861}

\bibitem[\protect\citeauthoryear{{Gonz\'alez-Alfonso}
  et~al.,}{{Gonz\'alez-Alfonso} et~al.}{2017}]{WP3-Gonzales-Alfonso}
{Gonz\'alez-Alfonso} E.,  et~al., 2017, \pasa, this issue

\bibitem[\protect\citeauthoryear{{Gruppioni} et~al.,}{{Gruppioni}
  et~al.}{2017}]{WP4-Gruppioni}
{Gruppioni} C.,  et~al., 2017, \pasa, this issue

\bibitem[\protect\citeauthoryear{{Hijmering}, {den Hartog}, {Ridder}, {van der
  Linden}, {van der Kuur}, {Gao}  \& {Jackson}}{{Hijmering}
  et~al.}{2016}]{2016SPIE.9914E..1CH}
{Hijmering} R.~A.,  {den Hartog} R.,  {Ridder} M.,  {van der Linden} A.~J.,
  {van der Kuur} J.,  {Gao} J.~R.,   {Jackson} B.,  2016, in Millimeter,
  Submillimeter, and Far-Infrared Detectors and Instrumentation for Astronomy
  VIII. p. 99141C, \mndoi{10.1117/12.2231714}

\bibitem[\protect\citeauthoryear{{Kamp}}{{Kamp}}{2017}]{KampInprep}
{Kamp} I.,  2017, \aap, in prep

\bibitem[\protect\citeauthoryear{{Kaneda} et~al.,}{{Kaneda}
  et~al.}{2017}]{WP5-Kaneda}
{Kaneda} H.,  et~al., 2017, \pasa, this issue

\bibitem[\protect\citeauthoryear{{Kessler} et~al.,}{{Kessler}
  et~al.}{1996}]{1996A&A...315L..27K}
{Kessler} M.~F.,  et~al., 1996, \aap, \href
  {http://adsabs.harvard.edu/abs/1996A%26A...315L..27K} {315, L27}

\bibitem[\protect\citeauthoryear{{Khosropanah} et~al.,}{{Khosropanah}
  et~al.}{2016}]{2016SPIE.9914E..0BK}
{Khosropanah} P.,  et~al., 2016, in Millimeter, Submillimeter, and Far-Infrared
  Detectors and Instrumentation for Astronomy VIII. p. 99140B,
  \mndoi{10.1117/12.2233472}

\bibitem[\protect\citeauthoryear{{Klein} et~al.,}{{Klein}
  et~al.}{2014}]{2014SPIE.9147E..2XK}
{Klein} R.,  et~al., 2014, in Ground-based and Airborne Instrumentation for
  Astronomy V. p. 91472X, \mndoi{10.1117/12.2055371}

\bibitem[\protect\citeauthoryear{{Linder}, {Falkner}, {Puig}, {Renk}, {Doyle},
  {Timm}, {Symonds}  \& {Walker}}{{Linder} et~al.}{2014}]{Report:CDFCryoTel}
{Linder} M.,  {Falkner} P.,  {Puig} L.,  {Renk} F.,  {Doyle} D.,  {Timm} R.,
  {Symonds} K.,   {Walker} R.,  2014, {CDF Study Report CDF-152-A -
  NG-CryoIRTel Assessment of Next Generation Cryogenic Infra Red Telescope}

\bibitem[\protect\citeauthoryear{{Martin} \& {Puplett}}{{Martin} \&
  {Puplett}}{1970}]{MartinPuplett}
{Martin} D.,  {Puplett} E.,  1970, Infrared Physics, 10, 105

\bibitem[\protect\citeauthoryear{{McCarthy} et~al.,}{{McCarthy}
  et~al.}{2016}]{2016SPIE.9906E..12M}
{McCarthy} P.~J.,  et~al., 2016, in Ground-based and Airborne Telescopes VI. p.
  990612, \mndoi{10.1117/12.2234505}

\bibitem[\protect\citeauthoryear{{Mizutani}, {Yamawaki}, {Komatsu}, {Goto},
  {Takeuchi}, {Shinozaki}, {Matsuhara}  \& {Nakagawa}}{{Mizutani}
  et~al.}{2015}]{2015JATIS...1b7001M}
{Mizutani} T.,  {Yamawaki} T.,  {Komatsu} K.,  {Goto} K.,  {Takeuchi} S.,
  {Shinozaki} K.,  {Matsuhara} H.,   {Nakagawa} T.,  2015, \mndoi [Journal of
  Astronomical Telescopes, Instruments, and Systems]
  {10.1117/1.JATIS.1.2.027001}, \href
  {http://adsabs.harvard.edu/abs/2015JATIS...1b7001M} {1, 027001}

\bibitem[\protect\citeauthoryear{{Moorwood}}{{Moorwood}}{1999}]{1999ESASP.427..825M}
{Moorwood} A.~F.~M.,  1999, in {Cox} P.,  {Kessler} M.,  eds,  ESA Special
  Publication Vol. 427, The Universe as Seen by ISO. p.~825

\bibitem[\protect\citeauthoryear{{Murakami} et~al.,}{{Murakami}
  et~al.}{2007}]{2007PASJ...59S.369M}
{Murakami} H.,  et~al., 2007, \mndoi [\pasj] {10.1093/pasj/59.sp2.S369}, \href
  {http://adsabs.harvard.edu/abs/2007PASJ...59S.369M} {59, S369}

\bibitem[\protect\citeauthoryear{{Nakagawa} et~al.,}{{Nakagawa}
  et~al.}{1998}]{1998SPIE.3356..462N}
{Nakagawa} T.,  et~al., 1998, in {Bely} P.~Y.,  {Breckinridge} J.~B.,  eds,
  \procspie Vol. 3356, Space Telescopes and Instruments V. pp 462--470,
  \mndoi{10.1117/12.324469}

\bibitem[\protect\citeauthoryear{{Nakagawa}, {Shibai}, {Onaka}, {Matsuhara},
  {Kaneda}, {Kawakatsu}  \& {Roelfsema}}{{Nakagawa}
  et~al.}{2014}]{2014SPIE.9143E..1IN}
{Nakagawa} T.,  {Shibai} H.,  {Onaka} T.,  {Matsuhara} H.,  {Kaneda} H.,
  {Kawakatsu} Y.,   {Roelfsema} P.,  2014, in Space Telescopes and
  Instrumentation 2014: Optical, Infrared, and Millimeter Wave. p. 91431I,
  \mndoi{10.1117/12.2055947}

\bibitem[\protect\citeauthoryear{{Neugebauer} et~al.,}{{Neugebauer}
  et~al.}{1984}]{1984ApJ...278L...1N}
{Neugebauer} G.,  et~al., 1984, \mndoi [\apjl] {10.1086/184209}, \href
  {http://adsabs.harvard.edu/abs/1984ApJ...278L...1N} {278, L1}

\bibitem[\protect\citeauthoryear{{Notsu}, {Nomura}, {Ishimoto}, {Walsh},
  {Honda}, {Hirota}  \& {Millar}}{{Notsu} et~al.}{2016}]{2016ApJ...827..113N}
{Notsu} S.,  {Nomura} H.,  {Ishimoto} D.,  {Walsh} C.,  {Honda} M.,  {Hirota}
  T.,   {Millar} T.~J.,  2016, \mndoi [\apj] {10.3847/0004-637X/827/2/113},
  \href {http://adsabs.harvard.edu/abs/2016ApJ...827..113N} {827, 113}

\bibitem[\protect\citeauthoryear{{Notsu}, {Nomura}, {Ishimoto}, {Walsh},
  {Honda}, {Hirota}  \& {Millar}}{{Notsu} et~al.}{2017}]{2017ApJ...836..118N}
{Notsu} S.,  {Nomura} H.,  {Ishimoto} D.,  {Walsh} C.,  {Honda} M.,  {Hirota}
  T.,   {Millar} T.~J.,  2017, \mndoi [\apj] {10.3847/1538-4357/836/1/118},
  \href {http://adsabs.harvard.edu/abs/2017ApJ...836..118N} {836, 118}

\bibitem[\protect\citeauthoryear{{Ogawa} et~al.,}{{Ogawa}
  et~al.}{2016}]{2016SPIE.9904E..2HO}
{Ogawa} H.,  et~al., 2016, in Space Telescopes and Instrumentation 2016:
  Optical, Infrared, and Millimeter Wave. p. 99042H, \mndoi{10.1117/12.2231613}

\bibitem[\protect\citeauthoryear{{Pilbratt} et~al.,}{{Pilbratt}
  et~al.}{2010}]{2010A&A...518L...1P}
{Pilbratt} G.~L.,  et~al., 2010, \mndoi [\aap] {10.1051/0004-6361/201014759},
  \href {http://adsabs.harvard.edu/abs/2010A%26A...518L...1P} {518, L1}

\bibitem[\protect\citeauthoryear{{Planck Collaboration} et~al.,}{{Planck
  Collaboration} et~al.}{2011}]{2011A&A...536A...1P}
{Planck Collaboration} et~al., 2011, \mndoi [\aap]
  {10.1051/0004-6361/201116464}, \href
  {http://adsabs.harvard.edu/abs/2011A%26A...536A...1P} {536, A1}

\bibitem[\protect\citeauthoryear{{Poglitsch} et~al.,}{{Poglitsch}
  et~al.}{2010}]{2010A&A...518L...2P}
{Poglitsch} A.,  et~al., 2010, \mndoi [\aap] {10.1051/0004-6361/201014535},
  \href {http://adsabs.harvard.edu/abs/2010A%26A...518L...2P} {518, L2}

\bibitem[\protect\citeauthoryear{{Ridder}, {Khosropanah}, {Hijmering},
  {Suzuki}, {Bruijn}, {Hoevers}, {Gao}  \& {Zuiddam}}{{Ridder}
  et~al.}{2016}]{2016JLTP..184...60R}
{Ridder} M.~L.,  {Khosropanah} P.,  {Hijmering} R.~A.,  {Suzuki} T.,  {Bruijn}
  M.~P.,  {Hoevers} H.~F.~C.,  {Gao} J.~R.,   {Zuiddam} M.~R.,  2016, \mndoi
  [Journal of Low Temperature Physics] {10.1007/s10909-015-1381-z}, \href
  {http://adsabs.harvard.edu/abs/2016JLTP..184...60R} {184, 60}

\bibitem[\protect\citeauthoryear{{Rieke}, {Wright}, {Glasse}, {Ressler}  \&
  {MIRI Science Team}}{{Rieke} et~al.}{2010}]{2010AAS...21543904R}
{Rieke} G.,  {Wright} G.,  {Glasse} A.,  {Ressler} M.,   {MIRI Science Team}
  2010, in American Astronomical Society Meeting Abstracts \#215. p.~395

\bibitem[\protect\citeauthoryear{{Shinozaki} et~al.,}{{Shinozaki}
  et~al.}{2014}]{2014Cryo...64..228S}
{Shinozaki} K.,  et~al., 2014, \mndoi [Cryogenics]
  {10.1016/j.cryogenics.2014.03.011}, \href
  {http://adsabs.harvard.edu/abs/2014Cryo...64..228S} {64, 228}

\bibitem[\protect\citeauthoryear{{Shinozaki} et~al.,}{{Shinozaki}
  et~al.}{2016}]{2016SPIE.9904E..3WS}
{Shinozaki} K.,  et~al., 2016, in Space Telescopes and Instrumentation 2016:
  Optical, Infrared, and Millimeter Wave. p. 99043W, \mndoi{10.1117/12.2232602}

\bibitem[\protect\citeauthoryear{{Spinoglio} \& {Malkan}}{{Spinoglio} \&
  {Malkan}}{1992}]{1992ApJ...399..504S}
{Spinoglio} L.,  {Malkan} M.~A.,  1992, \mndoi [\apj] {10.1086/171943}, \href
  {http://adsabs.harvard.edu/abs/1992ApJ...399..504S} {399, 504}

\bibitem[\protect\citeauthoryear{{Spinoglio} et~al.,}{{Spinoglio}
  et~al.}{2017}]{WP1-Spinoglio}
{Spinoglio} L.,  et~al., 2017, \pasa, this issue

\bibitem[\protect\citeauthoryear{{Sugita} et~al.,}{{Sugita}
  et~al.}{2010}]{2010Cryo...50..566S}
{Sugita} H.,  et~al., 2010, \mndoi [Cryogenics]
  {10.1016/j.cryogenics.2010.02.026}, \href
  {http://adsabs.harvard.edu/abs/2010Cryo...50..566S} {50, 566}

\bibitem[\protect\citeauthoryear{{Suzuki} et~al.,}{{Suzuki}
  et~al.}{2016}]{2016JLTP..184...52S}
{Suzuki} T.,  et~al., 2016, \mndoi [Journal of Low Temperature Physics]
  {10.1007/s10909-015-1401-z}, \href
  {http://adsabs.harvard.edu/abs/2016JLTP..184...52S} {184, 52}

\bibitem[\protect\citeauthoryear{{Swinyard} et~al.,}{{Swinyard}
  et~al.}{2009}]{2009ExA....23..193S}
{Swinyard} B.,  et~al., 2009, \mndoi [Experimental Astronomy]
  {10.1007/s10686-008-9090-0}, \href
  {http://adsabs.harvard.edu/abs/2009ExA....23..193S} {23, 193}

\bibitem[\protect\citeauthoryear{{Tamai}, {Cirasuolo}, {Gonz{\'a}lez},
  {Koehler}  \& {Tuti}}{{Tamai} et~al.}{2016}]{2016SPIE.9906E..0WT}
{Tamai} R.,  {Cirasuolo} M.,  {Gonz{\'a}lez} J.~C.,  {Koehler} B.,   {Tuti} M.,
   2016, in Ground-based and Airborne Telescopes VI. p. 99060W,
  \mndoi{10.1117/12.2232690}

\bibitem[\protect\citeauthoryear{{Trapman}, {Miotello}, {Kama}, {van Dishoeck}
  \& {Bruderer}}{{Trapman} et~al.}{2017}]{2017A&A...605A..69T}
{Trapman} L.,  {Miotello} A.,  {Kama} M.,  {van Dishoeck} E.~F.,   {Bruderer}
  S.,  2017, \mndoi [\aap] {10.1051/0004-6361/201630308}, \href
  {http://adsabs.harvard.edu/abs/2017A%26A...605A..69T} {605, A69}

\bibitem[\protect\citeauthoryear{{Werner} et~al.,}{{Werner}
  et~al.}{2004}]{2004ApJS..154....1W}
{Werner} M.~W.,  et~al., 2004, \mndoi [\apjs] {10.1086/422992}, \href
  {http://adsabs.harvard.edu/abs/2004ApJS..154....1W} {154, 1}

\bibitem[\protect\citeauthoryear{{Wilson}, {Shopbell}, {Simpson},
  {Storchi-Bergmann}, {Barbosa}, {Ward}  \& {NASA/ESA}}{{Wilson}
  et~al.}{2000}]{Image:Wilson}
{Wilson} A.,  {Shopbell} P.,  {Simpson} C.,  {Storchi-Bergmann} T.,  {Barbosa}
  F. K.~B.,  {Ward} M.,   {NASA/ESA} 2000

\bibitem[\protect\citeauthoryear{{Wootten} \& {Thompson}}{{Wootten} \&
  {Thompson}}{2009}]{2009IEEEP..97.1463W}
{Wootten} A.,  {Thompson} A.~R.,  2009, \mndoi [IEEE Proceedings]
  {10.1109/JPROC.2009.2020572}, \href
  {http://adsabs.harvard.edu/abs/2009IEEEP..97.1463W} {97, 1463}

\bibitem[\protect\citeauthoryear{{Wright} et~al.,}{{Wright}
  et~al.}{2010}]{2010AJ....140.1868W}
{Wright} E.~L.,  et~al., 2010, \mndoi [\aj] {10.1088/0004-6256/140/6/1868},
  \href {http://adsabs.harvard.edu/abs/2010AJ....140.1868W} {140, 1868}

\bibitem[\protect\citeauthoryear{{van der Tak} et~al.,}{{van der Tak}
  et~al.}{2017}]{WP7-vanderTak}
{van der Tak} F.,  et~al., 2017, \pasa, this issue

\makeatother
\end{thebibliography}

\end{document}